\documentclass[aps,prd,onecolumn,groupedaddress]{revtex4-1}
\usepackage{graphicx}
\usepackage{color}

\newcommand{\beq}{\begin{equation}}
\newcommand{\eeq}{\end{equation}}
\newcommand{\bea}{\begin{eqnarray}}
\newcommand{\eea}{\end{eqnarray}}

\newcommand\pathint{\int {\cal D}U~}

\newcommand\trt{\frac14\textrm{tr}}
\newcommand\avr[1]{\left\langle{#1}\right\rangle}
\begin{document} 
\title{Fluctuations and correlations in high temperature QCD}

\author{
R. Bellwied$^{5}$,
S. Borsanyi$^{1}$,
Z. Fodor$^{1,2,3}$,
S. D. Katz$^{2,4}$,
A. P\'aszor$^{1}$,
C. Ratti$^{5}$,
K. K. Szabo$^{1,3}$
}
\affiliation{$^1$ \small{\it Department of Physics, Wuppertal University, Gaussstr. 20, D-42119 Wuppertal, Germany}\\
$^2$ \small{\it Inst. for Theoretical Physics, E\"otv\"os University,}\\
\small{\it P\'azm\'any P. s\'et\'any 1/A, H-1117 Budapest, Hungary}\\
$^3$ \small{\it J\"ulich Supercomputing Centre, Forschungszentrum J\"ulich, D-52425
J\"ulich, Germany}\\
$^4$ \small{\it MTA-ELTE "Lend\"ulet" Lattice Gauge Theory Research Group,}\\
\small{\it P\'azm\'any P. s\'et\'any 1/A, H-1117 Budapest, Hungary}\\
$^5$ \small{\it Department of Physics, University of Houston, Houston, TX 77204, USA}\\
}

\begin{abstract}
We calculate second- and fourth-order cumulants of conserved charges
in a temperature range stretching from the QCD transition region towards the
realm of (resummed) perturbation theory. We perform lattice simulations
with staggered quarks; the continuum extrapolation is based on
$N_t=10\dots24$ in the crossover-region and $N_t=8\dots16$ at higher
temperatures. We find that the Hadron Resonance Gas model predictions
describe the lattice data rather well in the confined phase. At high
temperatures (above $\sim$250 MeV) we find agreement with the three-loop 
Hard Thermal Loop results.
\end{abstract}


\maketitle

\section{Introduction\label{sec:intro}}

The Quark Gluon Plasma was formed in the Early Universe just a few microseconds after the Big Bang; 
today it is produced in heavy ion collision experiments at
the Large Hadron Collider (LHC) at CERN and the Relativistic Heavy Ion
Collider (RHIC) at Brookhaven Lab (BNL). This phase exists at high temperatures and/or densities, and is separated from the hadronic phase of Quantum Chromodynamics (QCD) by a cross-over transition \cite{Aoki:2006we}.
Lattice QCD has determined the temperature of this cross-over 
in Refs.~\cite{Aoki:2006br,Aoki:2009sc,Borsanyi:2010bp,Bazavov:2011nk}.

Below the transition temperature, the thermodynamics is governed by
massive hadrons with integer charges, whereas at high temperature nearly
free and nearly massless quarks with fractional charges and gluons dominate.
Fluctuations of various conserved charges are sensitive probes of the
quantum numbers and the associated masses, and have been proposed as a signal
of the deconfinement transition \cite{Jeon:2000wg,Asakawa:2000wh}.
In heavy ion experiments there is an ongoing effort
to measure the moments of conserved charge distributions \cite{Luo:2015cea}, which can be related one-to-one to fluctuations. They are particularly interesting for the beam energy
scan program at RHIC, since they may signal a nearby critical end point:
higher order moments of net proton distributions 
are sensitive to an increase in the correlation length \cite{Stephanov:1999zu}.
Fluctuations can also be used to extract the chemical freeze-out temperature
and chemical potential \cite{Karsch:2010ck}, as an alternative method
to the thermal fits to particle yields or ratios \cite{Becattini:2005xt,Cleymans:2005xv,Andronic:2005yp,Andronic:2008gu,Becattini:2014hla}.
The STAR collaboration has published the first four moments of the net-proton
\cite{Adamczyk:2013dal} and net-electric charge \cite{Adamczyk:2014fia}
distributions. In parallel to the experimental efforts,
the past years have witnessed a rapid development in the lattice
calculations of fluctuations \cite{Bazavov:2012vg,Borsanyi:2013hza},
leading to quantitative estimates of the chemical freeze-out temperature
and chemical potential for a range of RHIC energies \cite{Borsanyi:2014ewa}.

Diagonal quark number susceptibilities have already been
calculated in the early dynamical simulations \cite{Gottlieb:1987ac,Gottlieb:1988cq,Gavai:1989ce}; these were later complemented by the off-diagonal ones
\cite{Gavai:2001ie,Gavai:2002jt,Allton:2002zi,Bernard:2004je,Gavai:2005yk,Allton:2005gk}.
In the following years, higher order fluctuations have been calculated up to the sixth order
\cite{Allton:2005gk,Datta:2012pj},
with the main motivation to extrapolate several thermodynamic observables
to larger values of the chemical potential.
These were staggered simulations projects, but studies with Wilson quarks
are also emerging \cite{Borsanyi:2012uq,Borsanyi:2015waa,
Giudice:2013fza,Aarts:2014nba,Gattringer:2014hra}.
Strangeness fluctuations were used also to locate the transition temperature
and, for this purpose, they were continuum extrapolated.
With Wilson quarks this  was done with pion masses down to 285 MeV 
\cite{Borsanyi:2012uq,Borsanyi:2015waa}, for staggered quarks the continuum
limit was calculated at the physical point 
\cite{Aoki:2006br,Aoki:2009sc,Borsanyi:2010bp,Bazavov:2012jq}.
Continuum results for the other second cumulants appeared first in
Ref.~\cite{Borsanyi:2011sw} then in Ref.~\cite{Bazavov:2012jq}. Selected higher
order fluctuations were continuum extrapolated first in
Refs.~\cite{Borsanyi:2012rr,Borsanyi:2013hza,Bellwied:2013cta,Bazavov:2013uja}.

Below the cross-over temperature, hadrons (mesons and baryons) dominate the
thermodynamics. In this regime, the Hadron Resonance Gas (HRG) model provides a simple
description of thermodynamic quantities, including specific fluctuations or correlations \cite{Dashen:1969ep,Venugopalan:1992hy}.
Even before simulations with physical quark masses could be performed,
lattice QCD data were well described by the HRG model if the actual particle spectrum
was replaced by the unphysical one simulated on the lattice \cite{Karsch:2003zq,Huovinen:2009yb}.
The success of the HRG model based on the experimental resonance table has been demonstrated
later in several papers with physical quark masses and continuum extrapolation
for the chiral condensate \cite{Borsanyi:2010bp},
the equation of state \cite{Borsanyi:2013bia} 
and fluctuations \cite{Borsanyi:2011sw,Bazavov:2012jq}.

The concept of HRG has motivated new studies where
fluctuation-based observables were proposed for which, within the framework
of the HRG model, only particles and resonances with a specific quantum number contribute
(e.g. baryons in a specific strangeness sector) \cite{Bazavov:2013dta}. Since
at low $T$ most lattice results agree with the HRG predictions,
which is no longer true in the deconfined phase, the highest temperature of agreement can be
a model-dependent indicator of deconfinement, that can be
studied on a flavor-specific basis \cite{Bellwied:2013cta}.


Very high temperature QCD is best discussed in terms of improved perturbation
theory. The QCD thermodynamic potential is known up to $\alpha^3\log(\alpha)$
order \cite{Kajantie:2002wa}. This result was later generalized to finite
chemical potentials \cite{Vuorinen:2003fs} and the quark number
susceptibilities were calculated to the same order \cite{Vuorinen:2002ue}. The
soft contributions to these unimproved perturbative results can be resummed via
dimensional reduction \cite{Blaizot:2003iq}.  This idea has been applied to the
four-loop perturbative quark number susceptibilities
\cite{Andersen:2012wr,Mogliacci:2013mca}.

The hard thermal loop perturbation theory reorganizes the perturbative series,
enhancing its convergence \cite{Braaten:1991gm,Andersen:2002ey}. Recently, the
next-to-next-to-leading order pressure and energy density were calculated for the
SU(3) theory, \cite{Andersen:2010ct}, dramatically improving the agreement with
lattice simulations \cite{Boyd:1996bx,Borsanyi:2012ve}. Soon afterwards, the full QCD result was
calculated, too \cite{Strickland:2010tm,Andersen:2011ug}.  Fluctuations
were calculated at one
\cite{Andersen:2012wr}, two \cite{Haque:2013qta} and three loop order
\cite{Haque:2013sja}, improving the earlier HTL calculations of
susceptibilities \cite{Blaizot:2001vr,Blaizot:2002xz}.  This result was later
generalized to finite chemical potentials \cite{Haque:2014rua}.

In general, these highly resummed perturbative results are expected to provide
a good approximation, but their range of validity can only be determined if they
are compared with a non-perturbative approach, e.g. lattice QCD simulations. Such comparisons
have already been made, first on the level of the equation of state 
\cite{Andersen:2011ug}. Unfortunately, for this observable, the renormalization
scale-dependence is too large for a definitive answer on the range of validity.
Fluctuations, however, offer a more strict test for these diagrammatic approaches
because of the rather small renormalization scale dependence of the result from
dimensional reduction (DR) \cite{Mogliacci:2013mca} and from hard thermal loops
(HTL) \cite{Haque:2013sja}. Today lattice calculations at high temperatures are available e.g. with
the HISQ action of the BNL-Bielefeld group
\cite{Bazavov:2012jq,Bazavov:2013uja} and also with the 2stout action of the
Wuppertal-Budapest collaboration \cite{Borsanyi:2011sw,Borsanyi:2012rr}. 

In this paper, we present results on diagonal and non-diagonal second and fourth order fluctuations, in a temperature range which stretches from the transition region to the perturbation theory domain. 
Our simulations are performed within the 2nd generation staggered thermodynamics program (4stout action). We start with the discussion of the conserved charges in the grand canonical field theory and provide details on how their fluctuations
are calculated on the lattice. After describing our lattice thermodynamics
program, the scale setting procedure and the finite temperature simulations,
we highlight the technical challenges of a continuum extrapolation and
the estimate of the systematic error on the continuum results.
The results are organized in two sections. First we consider the cross-over
region, around the point where the Hadron Resonance Gas loses its
predictive power. Afterwards we compare our data to
(resummed) perturbative results at high temperatures. We close with some
concluding remarks pointing to further directions of research.

\section{Fluctuations in lattice QCD\label{sec:fluct}}

\subsection{QCD as a grand canonical ensemble}

In a canonical ensemble, the conserved charges are external parameters.
In a heavy ion collision, for example, the number of baryons, their 
electric charge and the vanishing strangeness are fixed during the entire
collision, expansion of the plasma and freeze-out. A grand canonical
ensemble emerges if a small sub-system is considered, that is still large
enough to be close to the thermodynamic limit \cite{Bzdak:2012an}.

In QCD there exists a conserved charge for each quark flavor, thus one
can introduce four quark chemical potentials in a $2+1+1$ flavor
system: $\mu_u$, $\mu_d$, $\mu_s$ and $\mu_c$, in short $\{\mu_q\}$.

The expectation number of a conserved charge is then found as
a derivative with respect to the chemical potential.
\begin{equation}
\langle N_i\rangle =  T \frac{\partial\log Z(T,V,\{\mu_i\})}{\partial\mu_i}
\end{equation}

The response of the system to the thermodynamic force $\mu_i$ is proportional
to the fluctuation of the conserved charge:
\begin{equation}
\frac{\partial\langle N_i\rangle}{\partial\mu_j}=
T \frac{\partial^2 \log Z(T,V,\{\mu_q\})}{\partial\mu_j\partial\mu_i}
=
\frac{1}{T}(\langle N_i N_j \rangle -\langle N_i  \rangle \langle N_j \rangle)
\label{eq:fluctresponse}
\end{equation}

Since $N_i$ is an extensive thermodynamic quantity and so is its 
$\mu$-derivative, there the $O(V^2)$ contributions cancel in
Eq.~(\ref{eq:fluctresponse}).
Charge conjugation symmetry implies that, at $\mu_q\equiv0$, the expectation
value of any odd combination vanishes, e.g. the last term in 
Eq.~(\ref{eq:fluctresponse}). However, there is no such symmetry 
for different flavors, allowing e.g. for a $\langle N_u N_d\rangle$
correlator. The first perturbative diagram that contributes to the latter
consists of two fermion loops, connected by three gluon lines
\cite{Blaizot:2001vr}.

The free energy density ($-T/V \log Z$) is proportional to the pressure in
large volumes: \begin{equation}
\frac{p}{T^4}=\frac{1}{VT^3}\log Z(T,V,\{\mu_q\})\,.
\end{equation}
The derivatives with respect to the chemical potential can thus be written
in terms of the pressure:
\begin{equation}
\chi^{u,d,s,c}_{i,j,k,l}= \frac{\partial^{i+j+k+l} (p/T^4)}{
(\partial \hat\mu_u)^i
(\partial \hat\mu_d)^j
(\partial \hat\mu_s)^k
(\partial \hat\mu_c)^l
}
\end{equation}
with $\hat\mu_q=\mu_q/T$.  This normalization ensures that the cumulants
stay dimensionless, and become finite in the infinite volume and infinite
temperature limit. In this normalization $\chi_1(T,\{\mu_q\})$ 
is the expected number of quarks of the given flavor in a volume $T^{-3}$.

The higher derivatives with respect to the same quark chemical potential
correspond to the higher moments of that flavor:
\begin{eqnarray}
\mathrm{ mean:}~~M=\chi_1~~&&~~\mathrm{ variance:}~~\sigma^2=\chi_2
\nonumber \\
\mathrm{ skewness:}~~S=\chi_3/\chi_{2}^{3/2}
~~&&~~
\mathrm{kurtosis:}~~\kappa=\chi_4/\chi_{2}^{2}\,.
\end{eqnarray}

In experiment, the net-charge distribution moments are measured, each carrying
an unknown volume factor. A known caveat is the fluctuation of these
volumes themselves. The study of these goes beyond the scope of this
paper, see \cite{Skokov:2012ds,Alba:2015iva}. For a fixed volume, though,
the volume factor can be simply cancelled out by forming ratios of cumulants
of the same conserved charge:
\begin{eqnarray}
~S\sigma=\chi_3/\chi_{2}
\quad&;&\quad
\kappa\sigma^2=\chi_4/\chi_{2}\nonumber\\
M/\sigma^2=\chi_1/\chi_2
\quad&;&\quad
S\sigma^3/M=\chi_3/\chi_1\,.
\label{moments}
\end{eqnarray}

Phenomenological models and experiments usually work in the 
baryon number ($B$) - electric charge ($Q$) - strangeness ($S$) basis.
Since the charm quark plays a negligible role in the transition region
one can express these directions in the ${\mu}$ space as a three-dimensional
transformation:
\begin{eqnarray}
\mu_u&=&\frac13\mu_B+\frac23\mu_Q\,,\\
\mu_d&=&\frac13\mu_B-\frac13\mu_Q\,,\\
\mu_s&=&\frac13\mu_B-\frac13\mu_Q-\mu_S\,.
\end{eqnarray}

The fluctuations of the conserved charges ($B$, $Q$ and $S$) can then
be expressed in terms of the quark derivatives. 
In addition, the ($z$ component of the) light isospin is
often studied with $\mu_I=(\mu_u-\mu_d)$.
Assuming zero chemical
potential and degenerate $u$ and $d$ quarks on the lattice, 
several simplifications occur, and we have
\cite{Bernard:2004je,Bazavov:2012jq}:

\begin{eqnarray}
\chi_2^{B}&=&\frac19 \left[
2\chi_2^u+\chi^s_2+4\chi_{11}^{us}+2\chi_{11}^{ud}
\right]\,,\\
\chi_2^{Q}&=&\frac19 \left[
5\chi^u_2+\chi^s_2-2\chi_{11}^{us}-4\chi_{11}^{ud}
\right]\,,\\
\chi_2^I&=&\frac{1}{2}\left[\chi_2^u-\chi_{11}^{ud}\right]\,,\\
\chi_{11}^{BQ}&=&\frac19 \left[
\chi^u_2-\chi^s_2-\chi_{11}^{us}+\chi_{11}^{ud}
\right]\,,\\
\chi_{11}^{BS}&=&-\frac13 \left[
\chi^s_2+2\chi_{11}^{us}
\right]\,,\\
\chi_{11}^{QS}&=&\frac13 \left[
\chi^s_2-\chi_{11}^{us}
\right]\,.
\end{eqnarray}

Indeed, due to the $u\leftrightarrow d$ degeneracy the six second order
combinations in the $B$, $Q$, $S$ space can be expressed in terms of
four quark correlators. 
There are 15 fourth order correlators in the ($B,Q,S$) space that can
be expressed in terms of 9 fourth order quark-correlators. The kurtosis
of the baryon and the electric charge is given by the following correlators:
\begin{eqnarray}
\chi_4^B&=\frac1{81}&\left[
2\chi^u_4+\chi^s_4+6\chi_{22}^{ud}+12\chi_{22}^{us}
+8\chi_{13}^{us}+8\chi_{31}^{us}+8\chi_{31}^{ud}
+24\chi_{211}^{uds}+12\chi_{112}^{uds}
\right]\,,\\
\chi^Q_{4}&=\frac1{81}&\left[
17\chi^u_4+\chi^s_4+24\chi_{22}^{ud}+30\chi_{22}^{us}
-4\chi_{13}^{us}-28\chi_{31}^{us}-40\chi_{31}^{ud}
+24\chi_{211}^{uds}-24\chi_{112}^{uds}
\right]\,,
\end{eqnarray}
other, mixed derivatives can be calculated analogously.

At high temperature, fluctuations approach the Stefan-Boltzmann limit.
For an ideal gas, the pressure at finite chemical potential reads \cite{kapusta:book,Toimela:1982hv}
\begin{equation}
\frac{p}{T^4}= \frac{8\pi^2}{45} + \frac{7\pi^2}{60} N_f + 
\frac12 \sum_f \left( \frac{\mu_f^2}{T^2} + \frac{\mu_f^4}{2\pi^2T^4}\right)\,.
\end{equation}
For the second and fourth order fluctuations this means that in the high
temperature limit $\chi_2\to 1$ and $\chi_4\to 6/\pi^2$, and no
mixed derivatives survive.

\subsection{Fluctuations on the lattice}

The standard way to introduce the chemical potential on the lattice is
to modify the temporal links, like the $A_4$ component of a homogeneous
U(1) field \cite{Hasenfratz:1983ba}:
\begin{eqnarray}
U_4(\mu)=e^\mu U_4,&\quad&U_4^+(\mu)=e^{-\mu} U_4^+
\end{eqnarray}
The fermion matrix $M$ is built from the $\mu$-dependent links.
In the staggered formalism, which we will use in this paper, each fermion
flavor may carry an independent chemical potential. The fermion determinants express a single quark flavor.
\begin{equation}
Z=\pathint e^{-S_g}
(\det{M_u})^{1/4}
(\det{M_d})^{1/4}
(\det{M_s})^{1/4}
(\det{M_c})^{1/4}\,,
\end{equation}
where $S_g$ is the gauge action. To be specific, in this paper we use the
tree-level Symanzik improvement in $S_g$, however its form plays no role in the
fluctuation-related formulas.
The derivative of the staggered fermion matrix $M$ takes the following from:
\begin{eqnarray}
\frac{dM}{d\mu}\psi(x)&=&\frac12\eta_4(x)
\left[U_4(x)\psi(x+\hat 4)+U_4^+(x-\hat0)\psi(x-\hat 4)\right],\nonumber\\
\frac{d^2M}{d\mu^2}\psi(x)&=&\frac12\eta_0(x)
\left[U_4(x)\psi(x+\hat 4)-U_4^+(x-\hat0)\psi(x-\hat 4)\right];\nonumber
\end{eqnarray}
any higher odd derivative is equal to $dM/d\mu$, while any higher even
derivative is equal to $d^2M/d\mu^2$. $\eta_\nu(x)$ is the Kogut-Susskind 
phase factor.

For the fourth order $\mu$-derivative one has to evaluate the fourth
derivatives of $\det M$. These are traces of the fermion matrix that have to be
calculated for every generated finite temperature configuration \cite{Allton:2002zi}:
\begin{eqnarray}
A_j&=\frac{d}{d\mu_j} \log (\det M_j)^{1/4} = &\trt M_j^{-1} M_j'\,,\\
B_j&=\frac{d^2}{(d\mu_j)^2} \log(\det M_j)^{1/4} =&\trt\left(
M_j'' M_j^{-1}
-M_j' M_j^{-1} M_j' M_j^{-1}
\right)\,,\\
C_j&=\frac{d^3}{(d\mu_j)^3} \log(\det M_j)^{1/4} =&\trt\left(
M_j' M_j^{-1}
-3 M_j'' M_j^{-1} M_j' M_j^{-1}\right.\nonumber\\
&&
\left.
+2 M_j' M_j^{-1} M_j' M_j^{-1} M_j' M_j^{-1}
\right)\,,\\
D_j&=\frac{d^4}{(d\mu_j)^4} \log(\det M_j)^{1/4} = &\trt\left(
M_j'' M_j^{-1}
-4 M_j' M_j^{-1} M_j' M_j^{-1}
-3 M_j'' M_j^{-1} M_j'' M_j^{-1} \right.\nonumber\\
&&
\left.
+12 M_j'' M_j^{-1} M_j' M_j^{-1} M_j' M_j^{-1}\right.\nonumber\\
&&
\left.
-6 M_j' M_j^{-1} M_j' M_j^{-1} M_j' M_j^{-1}M_j' M_j^{-1}
\right)\,,
\end{eqnarray}

Using the simple notation $\partial_j$ for $\partial/\partial\mu_j$,
the derivatives can now be written for the full free energy:
\begin{equation}
\partial_i \log Z = \avr{A_i}\,.
\end{equation}
The derivative of the expectation value of any $X$ lattice observable
is obtained as
\begin{equation}
\partial_j \avr{X}=\avr{X A_j}-\avr{X}\avr{A_j}+\avr{\partial_j X}\,.
\label{eq:Xderivative}
\end{equation}
When we derive the higher order formulas (see also \cite{Allton:2002zi})
we assume non-zero chemical potential and use Eq.~\ref{eq:Xderivative}
recursively. Setting in the end $\mu=0$ we have, to second order,
\begin{equation}
\partial_i\partial_j \log Z=
\avr{A_iA_j}-\avr{A_i}\avr{A_j}+\delta_{ij}\avr{B_i}\,,
\label{eq:Zuu}
\end{equation}
and to fourth order, exploiting the degeneracy
between the light quark flavors:
\begin{eqnarray}
\partial_i^4 \log Z&=&
\avr{A_i^4}-3\avr{A_i^2}^2+3\left(\avr{B_i^2}-\avr{B_i}^2\right)\nonumber\\
&&+6\left(\avr{A_i^2B_i}-\avr{A_i^2}\avr{B_i}\right)+4\avr{A_iC_i}
+\avr{D_i}\,,\label{eq:Zuuuu}\\
\partial_u^3\partial_d\log Z&=&
\avr{A_u^4}-3\avr{A_u^2}^2\nonumber\\
&&+3\left(\avr{A_i^2B_i}-\avr{A_i^2}\avr{B_i}\right)+\avr{A_iC_i}\,\\
\partial_u^2\partial_d^2\log Z&=&
\avr{A_u^4}-3\avr{A_u^2}^2+\avr{B_u^2}-\avr{B_u}^2\nonumber\\
&&+2\left(\avr{A_i^2B_i}-\avr{A_i^2}\avr{B_i}\right)\,\\
\partial_u^2\partial_s^2\log Z&=&
\avr{A_u^2A_s^2}-2\avr{A_uA_s}^2-\avr{A_u^2}\avr{A_s^2}
+\avr{B_uB_s}-\avr{B_u}\avr{B_s}\nonumber\\
&&+\avr{A_u^2B_s}-\avr{A_u^2}\avr{B_s}
+\avr{A_s^2B_u}-\avr{A_s^2}\avr{B_u}\,\label{eq:Zuuss}\\
\partial_u^3\partial_s\log Z&=&
\avr{A_u^3A_s}-3\avr{A_u^2}\avr{A_uA_s}\nonumber\\
&&+3\left(\avr{A_uA_sB_u}-\avr{A_uA_s}\avr{B_u}\right)+\avr{A_sC_u}\,\\
\partial_u\partial_s^3\log Z&=&
\avr{A_uA_s^3}-3\avr{A_s^2}\avr{A_uA_s}\nonumber\\
&&+3\left(\avr{A_uA_sB_s}-\avr{A_uA_s}\avr{B_s}\right)+\avr{A_uC_s}\,\\
\partial_u\partial_d\partial^2_s\log Z&=&
\avr{A_u^2A_s^2}-2\avr{A_uA_s}^2-\avr{A_u^2}\avr{A_s^2}\nonumber\\
&&+\avr{A_u^2B_s}-\avr{A_u^2}\avr{B_s}\,\\
\partial_u^2\partial_d\partial_s\log Z&=&
\avr{A_u^3A_s}-3\avr{A_uA_s}\avr{A_u^2}\nonumber\\
&&+\avr{A_uA_sB_u}-\avr{A_uA_s}\avr{B_u}\,.
\end{eqnarray}

We follow the standard stochastic strategy to calculate the traces $A\dots D$,
 and
evaluate them with a large number of Gaussian random sources.
If one is only interested in up to the fourth derivative, five
calls to the linear solver $Mx=b$ are necessary for each random source.
Since the operator $D$ appears only in connected contributions,
we do not need it to high accuracy. $A$, on the other hand, appears in
the disconnected term with the most difficult cancellation, so it needs
to be evaluated more often. $A$ requires one solver, while $C$
requires three solvers. Thus, if we evaluate $D$ with $N$ sources, we evaluate the
$A$ operator $8N$ times and the $B$ and $C$ operators $4N$ times.

It was pointed out in \cite{Allton:2002zi} that, when products of
traces are calculated (e.g. $\langle AA\rangle \sim \chi^{ud}_2$),
the two (or more) operators in the product must be calculated with
different (or uncorrelated) random sources. For this reason, we always
use quartets of independent sources. We typically use $N=128$ quartets
in our analysis. Multi-right-hand-side solvers are
particularly useful in this context, since these typically achieve a higher
flop rate on many supercomputers, because the gauge fields do not have to be
loaded from the memory with each source \cite{Kaczmarek:2014mga}.

The numerical evaluation of these diagrams with multiple random sources
can be accelerated by various means. One observation was that e.g. the
$A$ operator can be split into two parts $A_0+\delta A$, where $A_0$
is the result of a truncated solver and $\delta A$ is the difference between the
truncated result and the full precision solution. The advantage is that
$\delta A$ can be evaluated with less sources, while the more noisy 
$A_0$ is cheaper to work with \cite{Bali:2009hu}.

\section{Lattice action and ensembles\label{sec:lattice}}

This work is part of the second generation thermodynamics program of the
Wuppertal-Budapest collaboration. We use the tree-level Symanzik gauge action
with 2+1+1 flavors of four times stout smeared staggered quarks
\cite{Morningstar:2003gk}, with the smearing parameter $\rho=0.125$.

\subsection{Zero temperature simulations and the line of constant physics}

An essential step, before thermodynamics runs can be started with a new
action, is the tuning of the mass parameters and the determination of
the scale or, in other words, the mass and coupling renormalization of the
theory for each lattice cut-off that the thermodynamics project intends to
use. In this project we use degenerate up and down quarks. For simplicity,
we do not tune the charm mass separately but accept the continuum extrapolated
quark mass ratio $m_c/m_s=11.85$ of Ref.~\cite{Davies:2009ih}.
The light and strange quark masses are obtained by tuning the following ratios to their physical values:
\begin{equation}
R_S^{\rm phys}=\frac{2m_K^2-m_\pi^2}{f_\pi^2}=27.65,\quad R_L^{\rm phys}=\frac{m_\pi}{f_\pi}=1.069\,
\label{eq:physical_point}
\end{equation}
where we use the isospin-averaged pion and
kaon masses ($m_\pi$ and $m_K$) \cite{Aoki:2013ldr}.
$f_\pi=130.41~\mathrm{MeV}$ (see Ref.~\cite{Rosner:2013ica}) is used to set the scale.

In this work we use the zero temperature lattice configurations
produced for the 4stout $T=0$ project \cite{durrfkfpi}.
In the lattice spacing range $a=0.188~\mathrm{fm}~\dots~0.077~\mathrm{fm}$
we simulate four or more ensembles for eight inverse bare couplings
$\beta=6/g^2$. The RHMC streams for the ensembles are typically $\sim 2000$
trajectories long after thermalization.
We parametrized these ensembles such that they
form a $\pm3\%$ bracket around the physical point, which is defined in 
Eq.~(\ref{eq:physical_point}).  The box size of these
zero temperature simulations was without exception $Lm_\pi\gtrsim4$.

\begin{figure}
\centerline{\includegraphics[width=\textwidth]{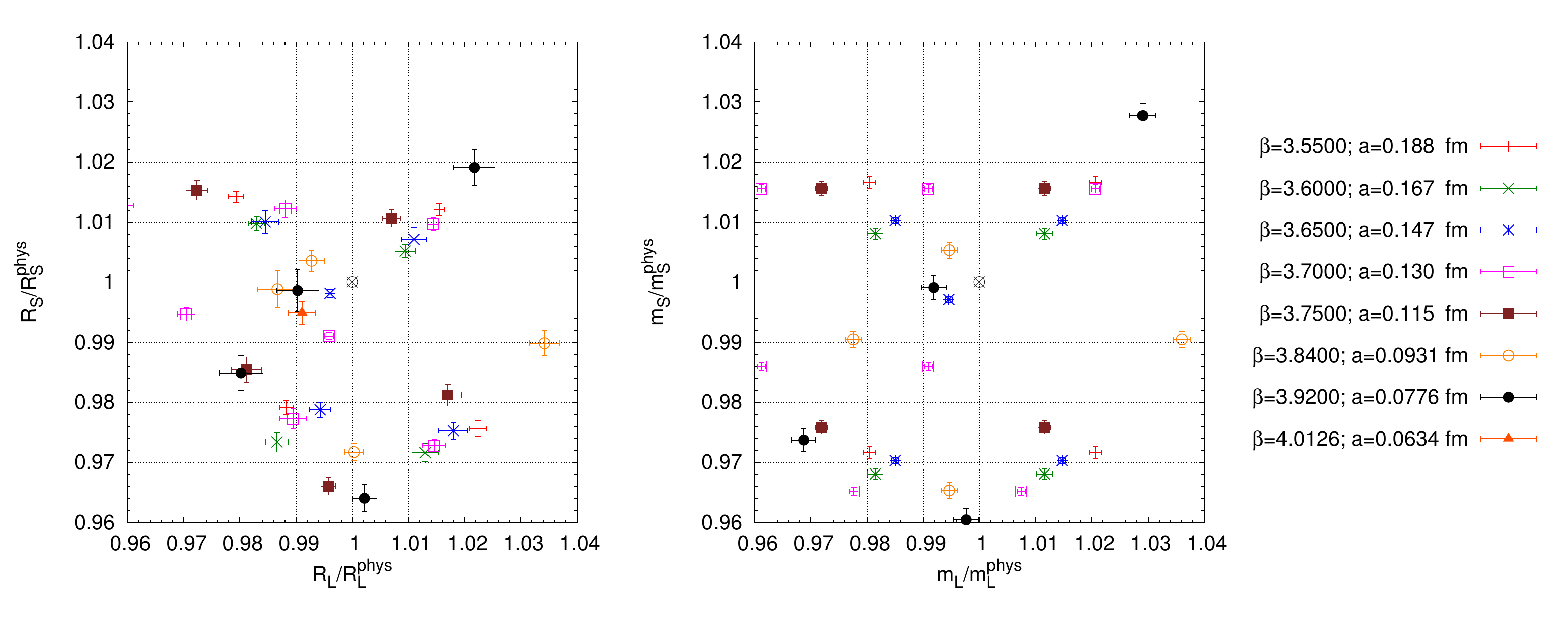}}
\caption{\label{fig:microlandscape}
Summary of our large volume zero temperature runs. The left panel
shows the combinations in Eq.~(\ref{eq:physical_point}) where
$R_S/R_S^{\rm phys}=1$ and $R_L/R_L^{\rm phys}=1$ define the physical
point. We determine the physical bare quark masses by interpolating
the bare parameters to precisely restore the $R_S$ and $R_L$ ratios
(see Ref.~\cite{durrfkfpi} for details).
The right panel shows the bare parameters rescaled by the interpolated
quark mass. The errors on the rescaled run parameters come solely from the
errors of the interpolation.
}
\end{figure}

In Fig.~\ref{fig:microlandscape} we summarize the zero temperature
configurations. For each $\beta$ we interpolated in the space of
bare quark masses, getting these to a few per mill accuracy.
On the left panel of Fig.~\ref{fig:microlandscape} we show the combinations in
Eq.~(\ref{eq:physical_point}). The right panel shows the position of individual
bare parameters relative to the thus interpolated physical point 
(with details given in Ref.~\cite{durrfkfpi}).

Our finest large volume ensemble was simulated at 
$\beta=4.0126$ on a $96^3\times144$ lattice. Its parameters were
extrapolated and then corrected using simulations at this $\beta$
in the flavor symmetric point, where all three light quark masses
are degenerate (the charm mass staying physical). 

\begin{figure}
\begin{center}
\includegraphics[width=3in]{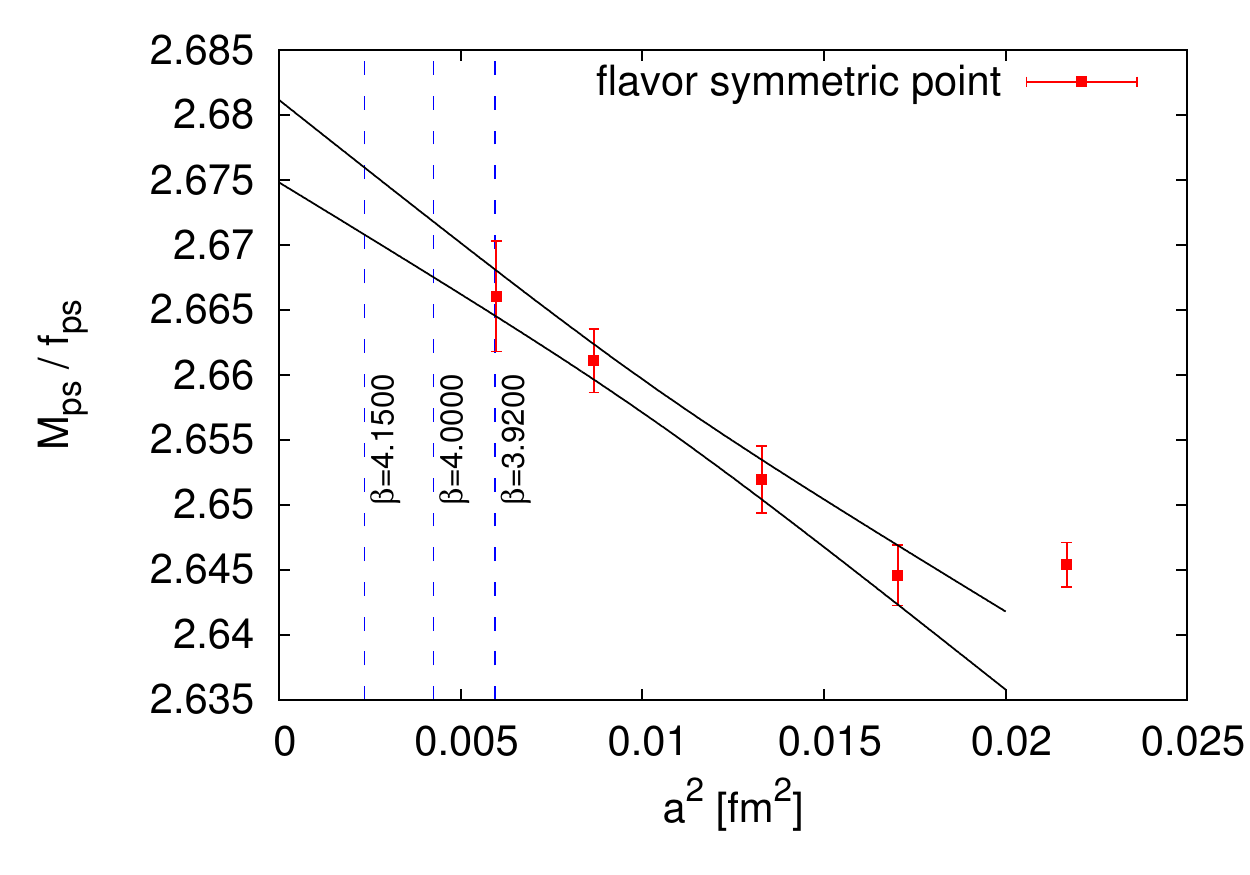}
\end{center}
\caption{\label{fig:fsp}
Using the bare parameters calculated from the 2+1+1 theory's quark masses
$\bar m=\frac13(m_u+m_d+m_s)$, $m_c=33.15 \bar m$, we find a mild $a^2$
dependence for the pseudo-scalar mass-to-decay-constant ratio in the 3+1
flavor (flavor symmetric) theory. The matching bare mass $\bar m$ at
larger $\beta$ (finer lattice) can be determined
at lower computational costs with 3+1 flavors. From $\bar m$, the bare masses of the
2+1+1 theory can be estimated.
}
\end{figure}

The tuning effort
using the flavor symmetric lattices goes as follows: first,
we have to acknowledge that various scale setting schemes
differ in the cut-off effects. Thus, changing the scale setting or tuning
principle may introduce different cut-off effects on different parts of the
line of constant physics. A continuum
extrapolation that spans a larger range of lattice spacings will thus
be distorted. To prevent this from happening, we match not only the scale
but also the $a^2$ corrections and check for the insignificance of the $a^4$ effects
whenever we are forced to switch between scale setting schemes along the line
of constant physics.  In this particular case,
we chose the mass-independent renormalization scheme. For a fixed gauge
coupling, we define a 3+1 flavor theory with the bare masses calculated
from the ones of the 2+1+1 flavor theory: $\bar m=\frac13(m_u+m_d+m_s)$,
 $m_c=33.15 \bar m$.  This corresponds to a new scheme, and the 
pseudo-scalar mass to decay constant ratio will have an $a^2$ dependence.
We plot this ratio in Fig.~\ref{fig:fsp}
(notice that, in the 2+1+1 theory, $m_\pi/f_\pi$ had no $a$-dependence by
definition). To extract the bare quark masses of the 2+1+1 dimensional
theory at $\beta=4.00$ and $\beta=4.15$, we performed
several simulations in the 3+1 flavor theory and interpolated 
$m_{\rm PS}/f_{\rm PS}$ in $\bar m$ to match the extrapolation in
Fig.~\ref{fig:fsp}.
We translated the masses back to the 2+1+1 flavor
theory. At this point, we had to assume the $m_s/m_u=27.63$ ratio, (which is consistent to our estimate from
this work) \cite{Davies:2009ih,McNeile:2010ji,
Durr:2010vn,Durr:2010aw}. For the large volume simulation at $\beta=4.0126$,
which was running with such an indirectly tuned mass, we show the result
in Fig.~\ref{fig:microlandscape}: the physical point is reproduced with an
accuracy below one percent. The lattice spacings are shown in the plot,
for the finest lattice we used the SU(2) low energy constants to extrapolate
the final one percent to the physical point \cite{Borsanyi:2012zv}.

\begin{figure}
\begin{center}
\includegraphics[width=3in]{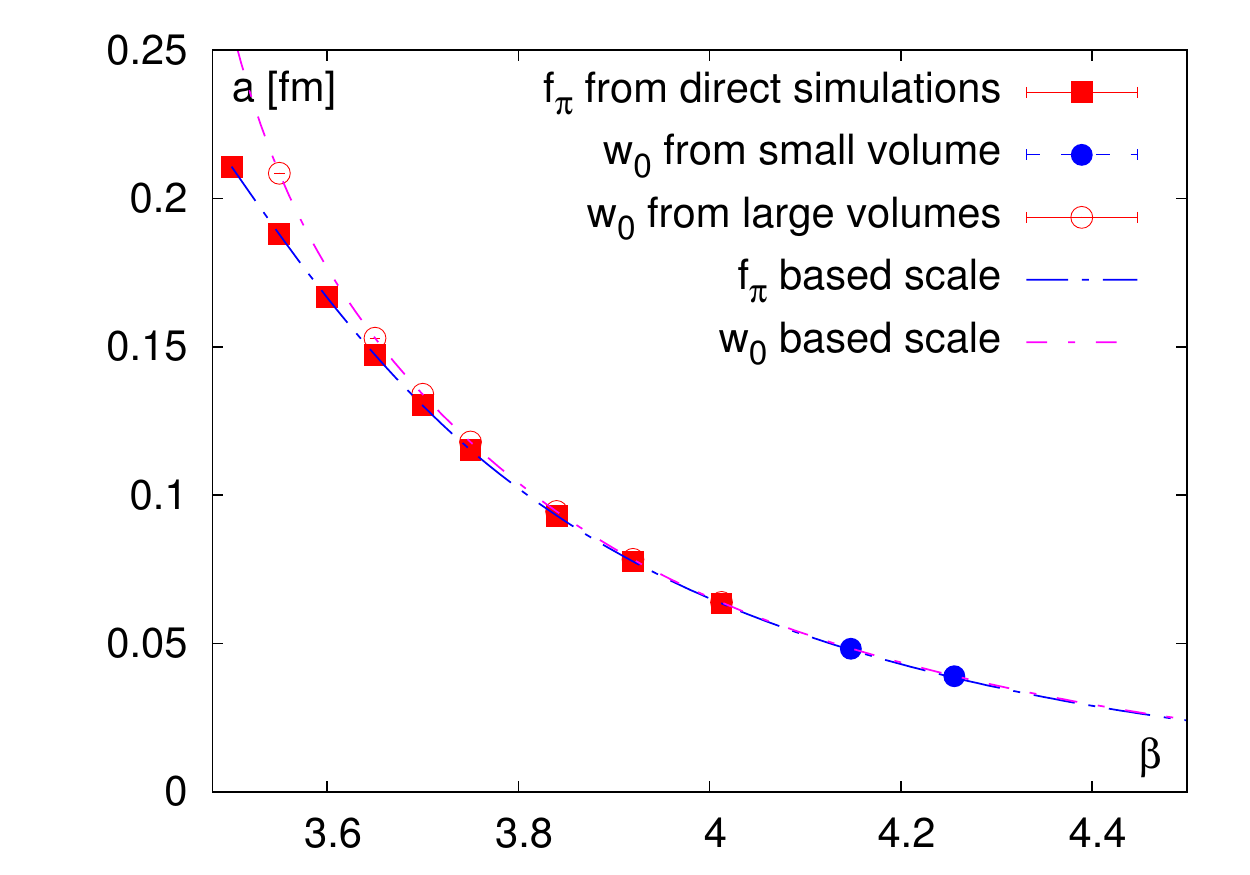}
\end{center}
\caption{\label{fig:scale}
The lattice spacing as a function of the inverse bare gauge coupling.
The red squares show the outcome of the zero-temperature simulations
with $Lm_\pi>4$ for $f_\pi$. The scale in the $w_0$ scheme from the same runs
is represented by the red circles. The blue dots correspond to smaller volumes, for which we used $w_0$ only. The differences coming from the two scale setting options
are part of our systematic error estimate.
}
\end{figure}

For even finer lattices we had to resort to a perturbative continuation
of the line of constant physics. For the scale setting, the universal
two-loop beta function does not yet describe the data. We have
an alternative scale setting scheme $w_0$, introduced in
\cite{Borsanyi:2012zs}, which is based on the gradient flow
\cite{Luscher:2010iy}. In that case, finite volume effects are small even for
lattices as small as $1.5$~fm \cite{Borsanyi:2012zs}. This allowed to match
again the value and $a^2$-dependence of $w_0$ at $\beta=4.1479$ ($a\approx
0.047$~fm) and $\beta=4.2562$ ($a\approx 0.038$~fm). The exploding
autocorrelation times have forced us to use extremely long update streams (cca.
50000 trajectories) in a $40^4$ volume. For even finer lattices we again
measured and matched the flow and its leading lattice artefacts in fixed
physical volume and topological sector in several subsequent steps.  The final
scale is plotted in Fig.~\ref{fig:scale}.  Since $w_0$ is of great interest for
a wider community we will discuss its value, volume-dependence and other
systematics in a publication devoted solely to scale setting.

Fig.~\ref{fig:scale} shows two versions of the scale setting. Controlled
continuum extrapolations are independent of the choice of the scale setting
scheme. The equivalence of the schemes on fine lattices is evident from
Fig.~\ref{fig:scale}. Nevertheless, this choice obviously influences the
temperature of a particular ensemble. 
 Especially for observables with large slope in temperature (e.g. the
quark number susceptibilities in the cross-over region) the scale setting
has an impact on the continuum scaling. We propagate this effect into
the final error bars by calculating the continuum limits with both
scale settings and include this in our study of systematics.

\subsection{Finite temperature ensembles}

We have generated three sets of ensembles, each with multiple lattice
spacings and temperatures. In the first set we use the aspect
ratio $LT=3$, which might have finite volume effects, but gives a more
favorable signal/noise ratio than larger volumes. The second set has
$LT=4$ and covers the entire transition range up to $2T_c$. Using
these ensembles we can conclude that, wherever it was possible to perform a
meaningful comparison (this includes all second order fluctuations
and cross-correlators), finite volume effects on the $LT=3$ ensembles
are negligible for any lattice spacing, let alone in the continuum limit which
is the largest source of systematic errors. We see significant finite volume
effects only in the chiral condensate and susceptibility, which are not part of
this study. For temperatures $T>300$~MeV we do not keep the lattice
geometry constant in our temperature scan, but keep the physical volume
more-or-less constant with $LT_c\gtrsim2$. For the finest,
$N_t=16$ lattices in this set we have thus used the lattices $80^3\times 16$,
$96^3\times 16$, $112^3\times 16$ and $128^3\times 16$ for $T=360,440,520$ and
600~MeV, respectively. In the high temperature range, the statistics
is limited to $\sim$1000 configuration / temperature / lattice spacing.

Table~\ref{tab:LT4} shows the statistics for the $LT=4$ ensembles in
the cross-over region and in the quark gluon plasma phase. The temperatures
below 150 MeV are used to compare the data to the predictions
of the Hadron Resonance Gas model. The $LT=3$ data set is restricted to
the cross-over region (see table~\ref{tab:LT3}). In the tables we give
the number of configurations that we have analyzed for generalized quark
number susceptibilities: these are separated by ten Rational Hybrid Monte Carlo
(RHMC) trajectories. The acceptance range varies between 80 and 95\%.

 In the absence of visible finite volume effects in this range, we combine the
results of these with the $LT=4$ data set to enhance the signal. Indeed, the
fluctuations of disconnected diagrams (especially $\avr{A^4}-3\avr{A^2}^2$) in
Eq.~(\ref{eq:Zuuuu}) are heavily penalized by large volumes. This contribution
also appears in the Taylor coefficients of the $\mu_B$ expansion
and is the main source of noise.

\begin{table}[h]
\begin{center}
\begin{tabular}{|c|c|c|c|c|c|}
\hline $T$ [MeV]&$32^3\times 8$&$40^3\times 10$&$48^3\times 12$&$64^3\times 16$&$80^3\times 20$\\\hline125& 10514 & 10080 & 10008 & 5027 & 2060 \\
130& 5766 & 4625 & 10253 & 5099 & 2000 \\
135& 14762 & 10590 & 10060 & 10189 & 2720 \\
140& 14863 & 5381 & 15043 & 4959 & 5097 \\
145& 5784 & 5020 & 10014 & 5019 & 1280 \\
150& 5464 & 5067 & 11043 & 5064 & 1631 \\
153& 4985 & 5517 & 6410 & 3641 & - \\
155& 5613 & 5001 & 10137 & 5015 & 1726 \\
157& 5526 & 5409 & 10018 & 5160 & 1065 \\
160& 5247 & 5017 & 4973 & 5073 & 1082 \\
165& 8169 & 10086 & 10496 & 5000 & 1000 \\
170& 6005 & 6113 & 5600 & 5111 &  600 \\
175& 12018 & 5375 & 5058 & 5104 &  972 \\
180& 5007 & 5089 & 5034 & 5013 & 1000 \\
190& 4900 & 5031 & 5121 & 5045 &  992 \\
200& 5989 & 5002 & 6722 & 1012 & 1000 \\
220& 5514 & 5000 & 7231 & 1003 & 1000 \\
240& 1712 & 5000 & 8082 & 3947 & 1000 \\
250& 10695 & 5685 & 5146 & - & - \\
260& 6287 & 5000 & 8623 & 5441 & 1000 \\
270& 11574 & 5682 & 5684 & - & - \\
280& 7067 & 5003 & 8751 & 1021 &  558 \\
290& 7316 & 5680 & 5684 & - & - \\
300& 5125 & 4917 & 5398 & 5310 & 1011 \\
\hline
\end{tabular}
\end{center}
\caption{\label{tab:LT4}
The statistics of lattices with $LT=4$ aspect ratio. The numbers count the
saved and analyzed configurations, each separated by ten RHMC
updates.
}
\end{table}

\begin{table}[h]
\begin{center}
\begin{tabular}{|c|c|c|c|c|c|}
\hline $T$ [MeV]&$24^3\times 8$&$32^3\times 10$&$40^3\times 12$&$48^3\times 16$&$64^3\times 24$\\
\hline
130& 39161 & 7736 & 10351 & - & 2007 \\
135& 41462 & 8724 & 10696 & 9892 & 3000 \\
140& 39867 & 8550 & 10240 & 8248 & 1551 \\
145& 40247 & 8518 & 10348 & 10130 & 2550 \\
150& 39996 & 8461 & 10569 & 6717 & 3044 \\
155& 19953 & 8625 & 10345 & 10211 & 1546 \\
160& 20015 & 9174 & 11611 & 10140 & 2063 \\
165& 10965 & 9750 & 10219 & 10136 & 1200 \\
\hline
\end{tabular}
\end{center}
\caption{\label{tab:LT3}
The number of analyzed configurations on lattices with $LT\approx 3$ aspect
ratio. The configurations are separated by ten RHMC updates.
}
\end{table}

\subsection{Continuum extrapolation}

The continuum extrapolation is mostly based on all available lattice
spacings. Since fine lattices have lower statistics, the coarsest $N_t=8$
results are usually included only in non-linear extrapolations, (e.g.
$A+B/N_t^2+C/N_t^4$ and other variations, where $A$ is the continuum limit).

While for some observables (e.g. $\chi^S_2(T)$, $\chi^B_2(T)$) 
there is a clear range of safe linear extrapolation (in most cases $N_t\ge10$),
observables that are related to pion physics (e.g. $\chi^Q_2(T)$,
$\chi^{ud}_{11}(T)$) show a very strong, non-linear $1/N_t^2$ dependence.
Only for very fine lattices ($N_t\gtrsim16$) we see a linear regime.
Such behaviour have been already reported for the second order
cumulants \cite{Borsanyi:2011sw,Bazavov:2012jq}.

\begin{figure}[t]
\begin{center}
\includegraphics[width=\textwidth]{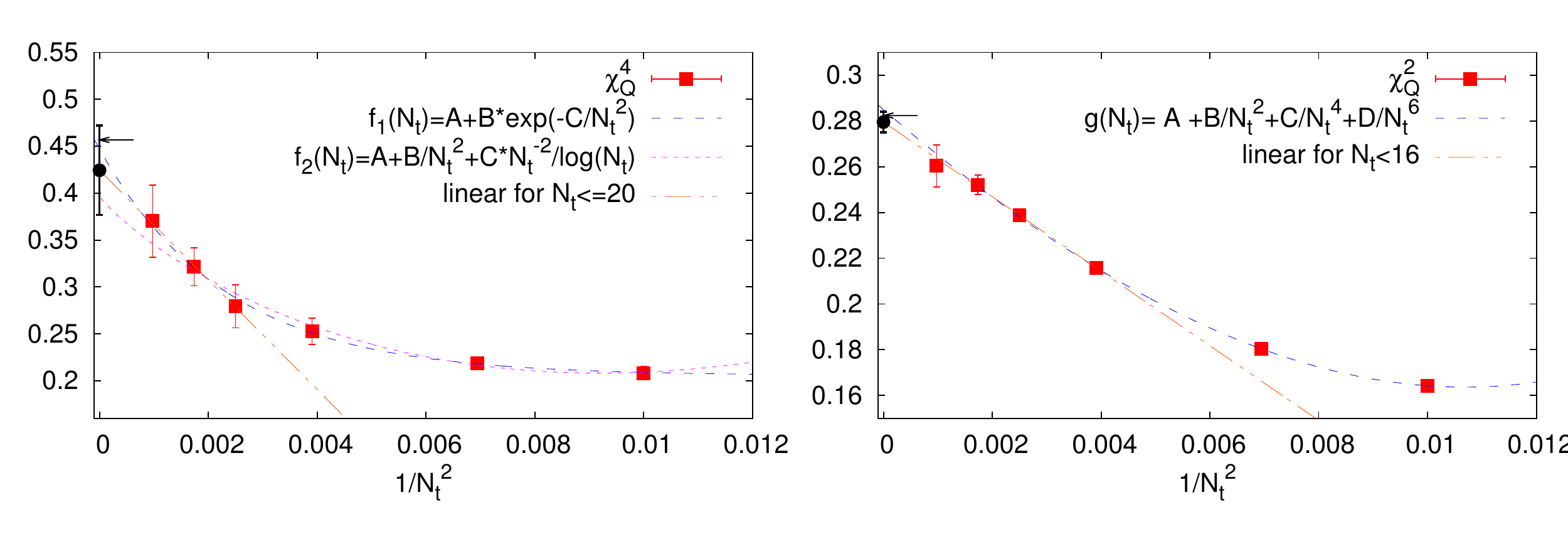}
\end{center}
\caption{\label{fig:T130_QQQQ}
Fluctuation data from the 4stout action at $T=130~\mathrm{MeV}$
for a lattice resolution range $N_t=10\dots32$. This corresponds to the lattice
spacings $a=0.15\dots0.047$~fm. A continuum extrapolation is possible by
fitting sophisticated models but also via a simple linear fit through the
finest lattices. The Hadron Resonance Gas model's prediction (small black
arrows) is consistent with the continuum limit.}
\end{figure}

Here we show the charge fourth and second moment for a single temperature
in the confined phase ($T=130$~MeV) in Fig.~\ref{fig:T130_QQQQ}. 
This plot features an additional $96^3\times32$ point with 1485 analyzed
configurations.
We attempt several fit models, $f_1(N_t)=A+B\exp(-C/N_t^2)$ 
resembles a Boltzmann factor with an artefact mass vanishing as 
$1/N_t^2$. $f_2(N_t)=A+B/N_t^2+C/N_t^2/\log(N_t)$ is similar to including a
$\alpha a^2$ term into the extrapolation. The shown continuum limit is based on
the linear fit only.

\begin{figure}
\centerline{
\hfil
\includegraphics[width=3in]{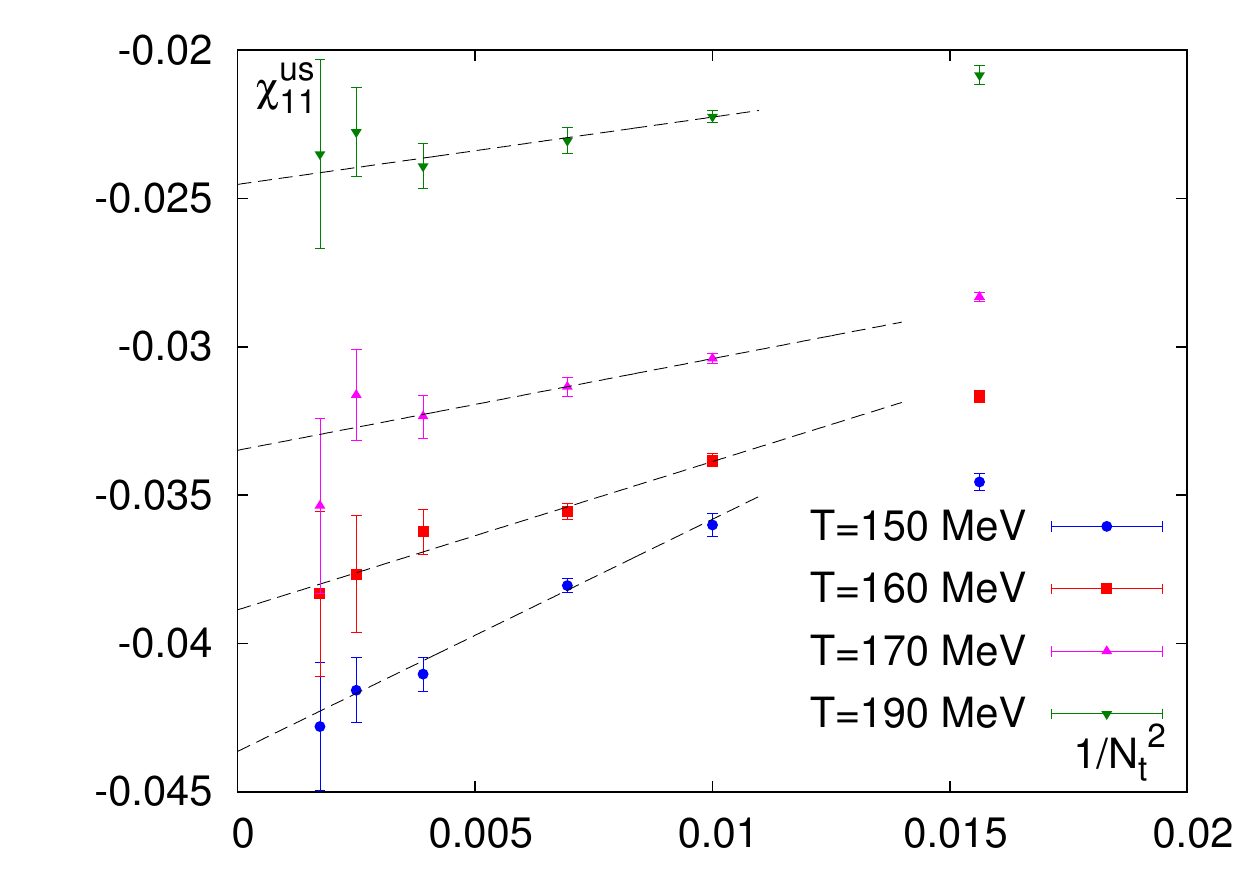}\hfil
\includegraphics[width=3in]{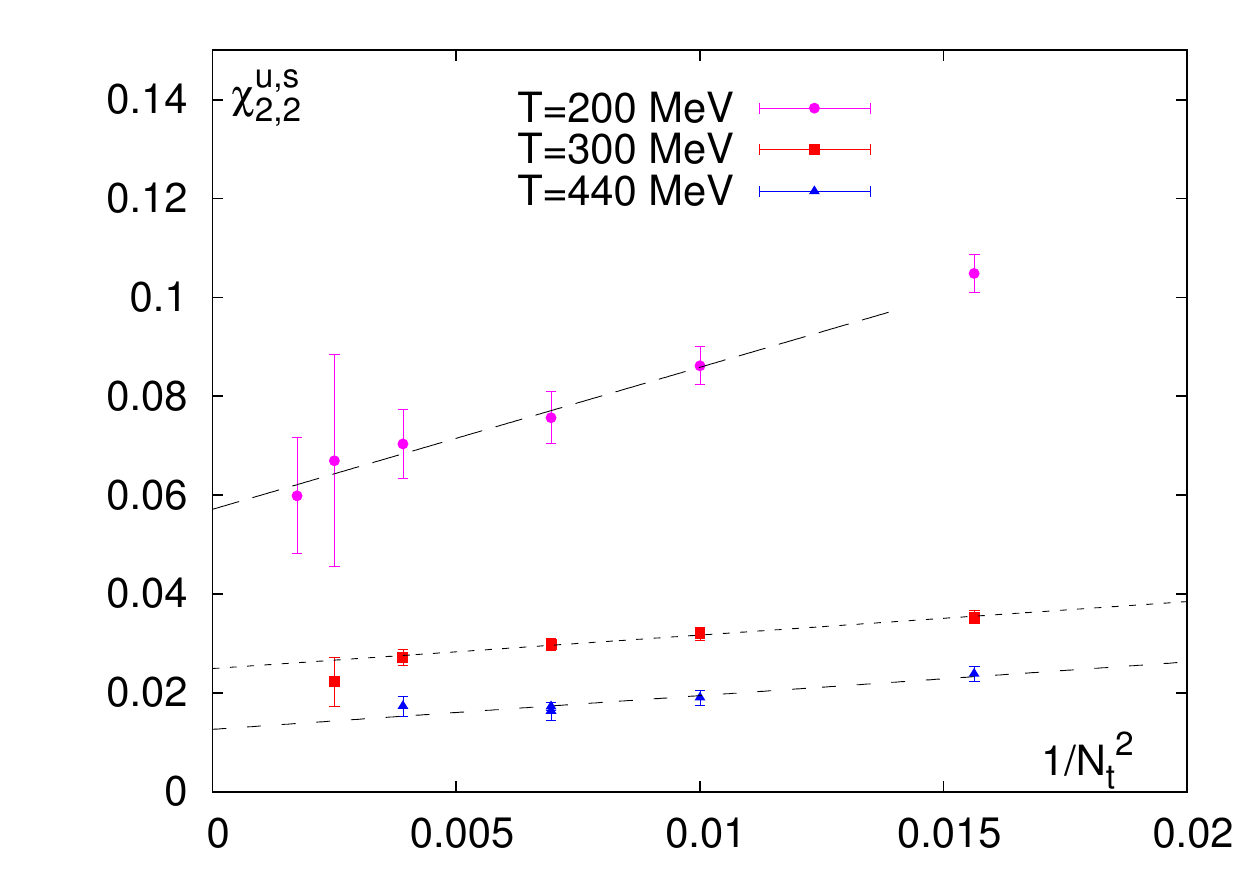}\hfil
}
\caption{\label{fig:c2us}
Examples of linear continuum extrapolations for the light-strange
correlators $\chi^{us}_{11}$ and $\chi^{us}_{22}$ at various temperatures.
The error on our continuum results contains the systematics of varying 
the fit model, fit window, scale setting and interpolation, see text.
}
\end{figure}

Not all observables require the finest lattices in our data set. Strange
quark correlators receive no pion contributions, and the small
relative taste violation in the kaon sector can be extrapolated away.
We find that our data with its current precision allow linear fitting for
$N_t\ge10$. As examples we show the up-strange correlator ($\chi^{us}_{11}$)
and the higher order correlator
between
the same quarks $\chi^{us}_{22}$ in Fig.~\ref{fig:c2us}.
 Both have only disconnected contributions
(see Eqs.~(\ref{eq:Zuu},\ref{eq:Zuuss})).

The parameters of the finite temperature runs have been tuned to have
the same temperature in the $f_\pi$ scale setting scheme. Since we
also use the $w_0$ scale setting scheme, in that case the temperatures are no
longer aligned and interpolations are necessary. The alignment of the
temperatures is also not perfect in the $f_\pi$ scale setting scheme, thus we
interpolate all data sets. The interpolation is performed by fitting a
spline through several (7-9) node points with two different sets of
nodes so that the systematics of the interpolation can be picked up by
the systematic error. We then perform the continuum extrapolation temperature
by temperature, for those temperatures for which we had data points. 
The lattice artefacts of the diagonal fluctuations can be understood
from tree-level perturbative diagrams \cite{Gavai:2002jt}. We can correct for
the $\alpha$-independent part of the discretization errors
by a $T$-independent factor (tree level improvement) \cite{Borsanyi:2010cj}.
This factor converges to 1 in the continuum limit. We perform 
the continuum extrapolation in three possible ways: without this improvement,
with the tree-level improvement, and with the improvement factor of the free
energy, that we find empirically to also reduce the cut-off effects at
intermediate temperatures.
 We must then judge for every
observable separately whether we can include the $N_t=8$ and $N_t=10$
ensembles, and which non-linear models are plausible and match the data.  We
have given examples for this in Fig.~\ref{fig:T130_QQQQ}, but very often we
simply add the models $A/(1+B/N_t^2)$ and $A+B/N_t^2+C/N_t^4$ to the linear
fit. 

We treat every mentioned option independently and perform 16--32 analyses per
temperature, depending on the complexity of the continuum scaling.
We use this large set of analyses to estimate the systematic errors
temperature by temperature using the histogram method introduced
in Refs.~\cite{Durr:2008zz,Borsanyi:2014jba}.
In this paper we build a histogram of the results. The analyses
with a fixed data set but different systematics are weighted using the Akaike
Information Criterion (AIC) \cite{Akaike}. The AIC weigted results corresponding
to the various fit windows in $1/N_t^2$ are combined with uniform weights.
In the case of the charm susceptibility we calculate the systematic errors
on the finite $N_t$ points first and then perform various continuum
extrapolations which then enter the histogram method.
Since all analyses are equally we identify the median with the result.
The distribution of results is not necessarily Gaussian and may contain
isolated combinations of the analysis options that produce outliers.
These do not contribute to the median. The systematic error is the
spread of the distribution. Instead of the standard deviation we use
the spread of central 68\% of the distribution, so that we do not have
to make assumptions on the tail of the distribution. The median
can be calculated for every jackknife or bootstrap sample. We use
the variance of the median as statistical error. 
In the plots we show the combined
errors, by adding up the systematic and statistical errors in quadrature.

\section{Results in the cross-over region\label{sec:lowT}}

Previous works have suggested that the Hadron Resonance Gas (HRG)
model provides a good description of the data in the range 
130-150 MeV \cite{Huovinen:2009yb,Borsanyi:2013bia, Borsanyi:2010bp,
Borsanyi:2011sw,Bazavov:2012jq}, and perhaps missing strange resonances might
account for the small deviations in the strangeness sector
\cite{Bazavov:2014xya}.

In this paper we supplement the picture with additional continuum extrapolated
data. Finite lattice spacing studies (with or without a well improved action)
can never state with certainty whether deviations from the model are a genuine effect.
Here we compare our lattice results using the 2014 edition of the
Particle Data Book \cite{Agashe:2014kda}.

In our previous paper \cite{Borsanyi:2011sw} we have calculated nearly all the
second order fluctuations. Only the most difficult correlator  was
omitted $\chi^{ud}_{11}(T)$, which is not only noisy but had severe
lattice spacing effects, similar to $\chi^Q_2(T)$ in Fig.~\ref{fig:T130_QQQQ}.

\begin{figure}[ht]
\begin{center}
\includegraphics[width=3in]{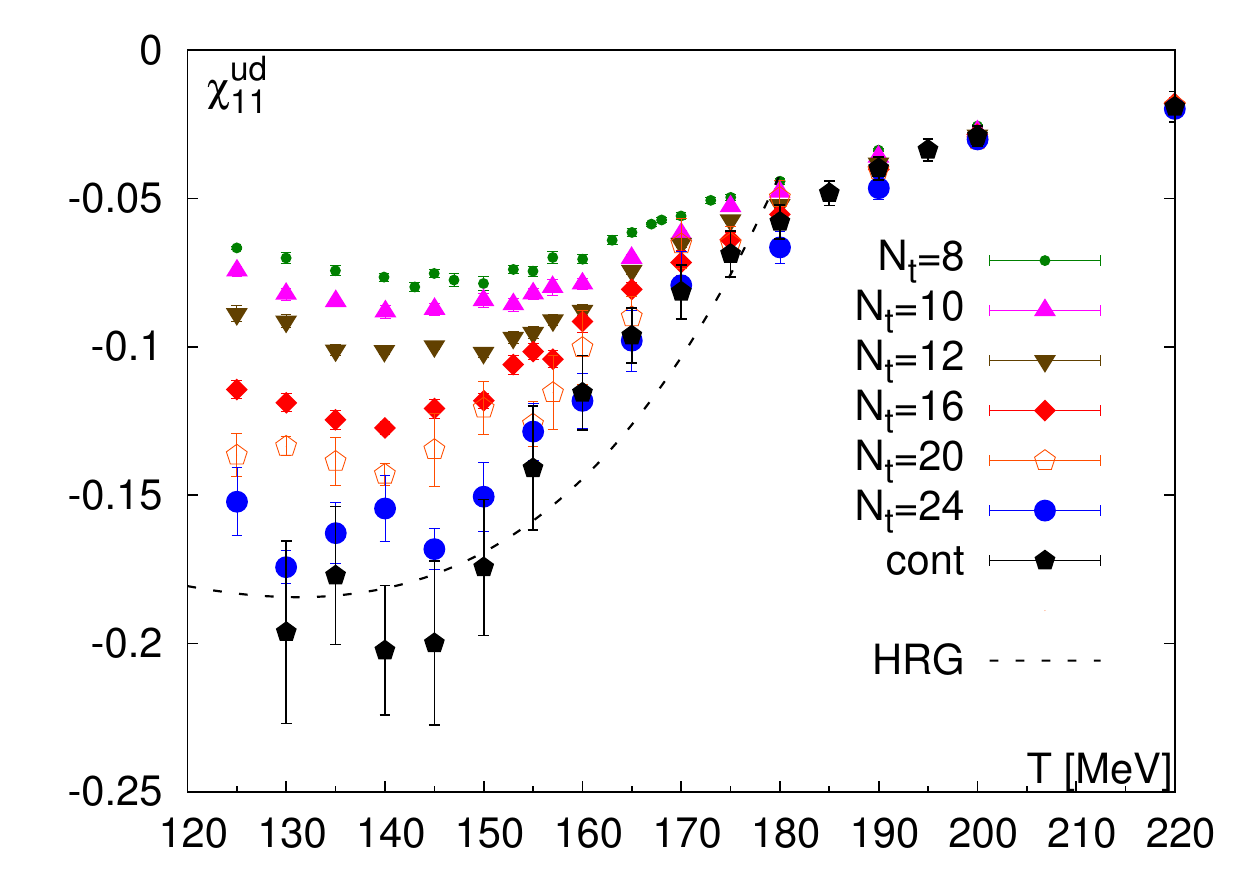}
\includegraphics[width=3in]{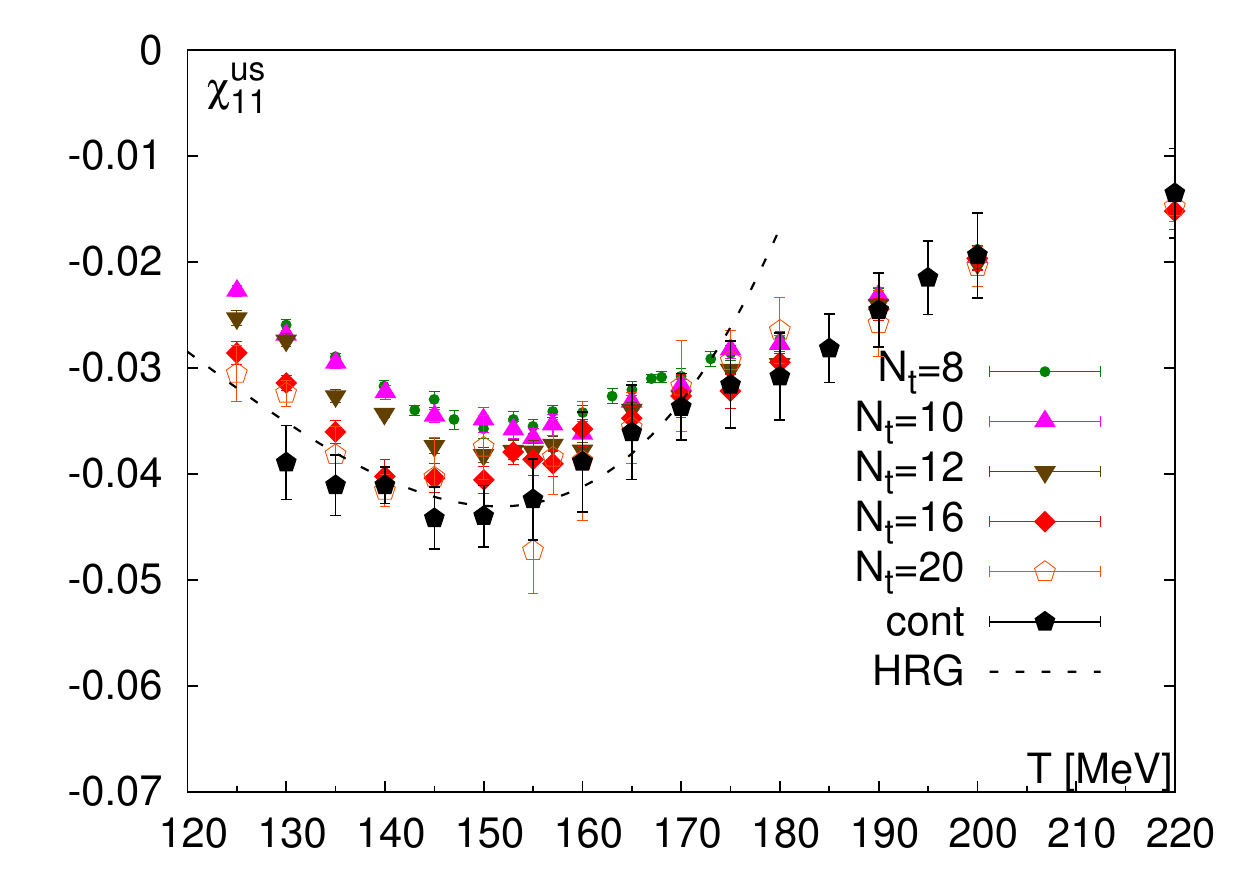}
\end{center}
\caption{\label{fig:c2ud}
The up-down correlator ($\chi^{ud}_{11}$) and up-strange correlator ($\chi^{us}_{11}$) for several lattice spacings and in the
continuum limit. For the Hadron Resonance Gas model we use the resonance
table in the 2014 edition of the Particle Data Book \cite{Agashe:2014kda}.
}
\end{figure}

The continuum extrapolation of $\chi^{ud}_{11}(T)$ and the data in the full
lattice spacing range are shown in Fig.~\ref{fig:c2ud}, together with the
up-strange correlator $\chi^{us}_{11}(T)$. The continuum limit
for $\chi^{ud}_{11}(T)$ is well described by the HRG model up to $T\approx
155$~MeV, which lies at the centre of the transition region
\cite{Borsanyi:2010bp,Bazavov:2011nk}.
The main hadrons that contribute to the HRG prediction are the light mesons,
mostly pions (the combination of a quark with an anti-quark makes the
$\chi^{ud}_{11}$ contribution negative). At high temperatures, heavier hadrons
and their resonances have non-negligible Boltzmann factors, allowing the
baryons (mostly protons) to take over the main role and bend the curve upwards.

The important role played by the pions is also highlighted by the staggered
lattice artefacts (taste breaking) that increase the mass of the various
staggered pion-like degrees of freedom (tastes) 
\cite{Bernard:2006ee}.

In lattice QCD $\chi^{ud}_{11}(T)\sim \avr{A_uA_d}/V$ in Eq.~(\ref{eq:Zuu})'s
notation. The $A=(1/4) \textrm{Tr} M^{-1}M'$ 
operator is a trace over the whole lattice. The normalized Gaussian 
random sources ($\chi$)  that we use to evaluate $A$, contribute each as
$\chi^+ M^{-1}M'\chi/4 \sim V$. This $C$-odd estimator is widely oscillating
between sources. Thus,
in $A$ and then also in the stochastic representation of $\avr{AA}/V$,
large cancellations occur between opposite-sign contributions. 
Refs.~\cite{Allton:2002zi,Ejiri:2008xt} link the phase of the
fermion determinant at small $\mu_B$ to the odd operators $A$ and $C$. Indeed,
the sign problem is already present in the Taylor-expansion technique and in
the calculation of baryonic fluctuations in general.

The consequence is that the severity of the sign problem is related to the magnitude
of $\chi^{ud}_{11}$. In early staggered studies one saw peak heights
of $\approx-0.005$ \cite{Allton:2002zi}, $\approx-0.014$ \cite{Allton:2005gk},
and $\approx-0.05$ \cite{Bernard:2004je}, are well short of today's continuum
limit in Fig.~\ref{fig:c2ud}. With the early actions and coarse lattices
the calculation of higher derivatives and reweighting were easier. 

Note that the light isospin susceptibility ($\chi^I_2$) does not depend on
the $A$ operator, $\chi^I_2\sim\avr{B}$, it does not contain any disconnected diagrams
at all. The fourth derivative $\chi^I_4\sim [6\avr{\delta B^2}-\avr{D}]/V$, too,
contains only $C$-even operators. Indeed, thermodynamics at finite 
isospin chemical potential is not plagued by the sign problem.

\begin{figure}[ht]
\begin{center}
\includegraphics[width=3in]{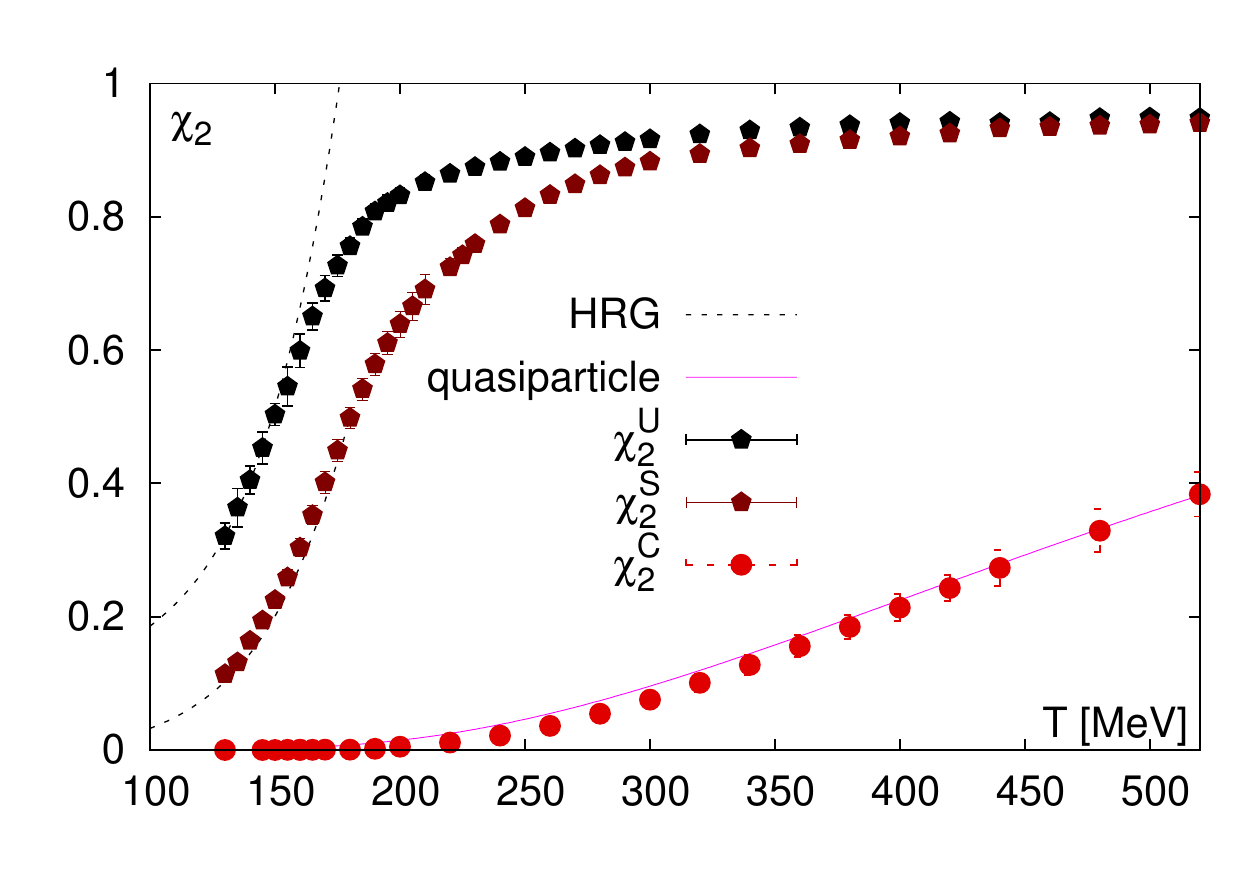}
\end{center}
\caption{\label{fig:c2cmp}
Light, strange and charm diagonal quark number susceptibilities in the continuum
limit as functions of the temperature. The quasi-particle model is calculated
for a single, non-interacting charm quark with a fitted mass
$m_c^{\mathrm{QP}}=1430~\mathrm{MeV}$.  }
\end{figure}

A subset of the authors of this paper have remarked that one can observe
a hierarchy between flavors in their fluctuations \cite{Bellwied:2013cta}.
We are now extending the picture and show the continuum extrapolations
of the flavor-specific quark number susceptibilities in Fig.~\ref{fig:c2cmp}.
The HRG model describes the light flavors reasonably well.
The charm susceptibility in Fig.~\ref{fig:c2cmp} rises at higher temperatures,
compared to the lighter flavors.  It was emphasized in
Ref.~\cite{Bazavov:2014yba} that open charm with fractional baryon charge
starts appearing near the chiral crossover temperature.
In addition to the hadron resonance gas model we show a naive quasiparticle
estimate for the charm susceptibility (see also \cite{Petreczky:2009cr}). The
mass of the charm quark was fitted to the last points
($m_c^{\mathrm{QP}}=1430~\mathrm{MeV}$). 
This mass is empirical, and may depend on the range of the matching to our
lattice data. In general the mass of the charm quark is scheme dependent.
The susceptibility curve runs near the
quasiparticle model, qualitatively confirming that $\chi^C_2$ is 
contributed to by the deconfined charm quark. Nevertheless,
the quasiparticle model's results are overestimating the lattice data below approx. 350 MeV.
This leaves room for multiple interpretations (e.g. $T$-dependent $m_c^{\mathrm{QP}}$,
limitations of the quasiparticle model
or charmonium bound states that absorb some of the free quarks).

\begin{figure}[ht]
\begin{center}
\includegraphics[width=3in]{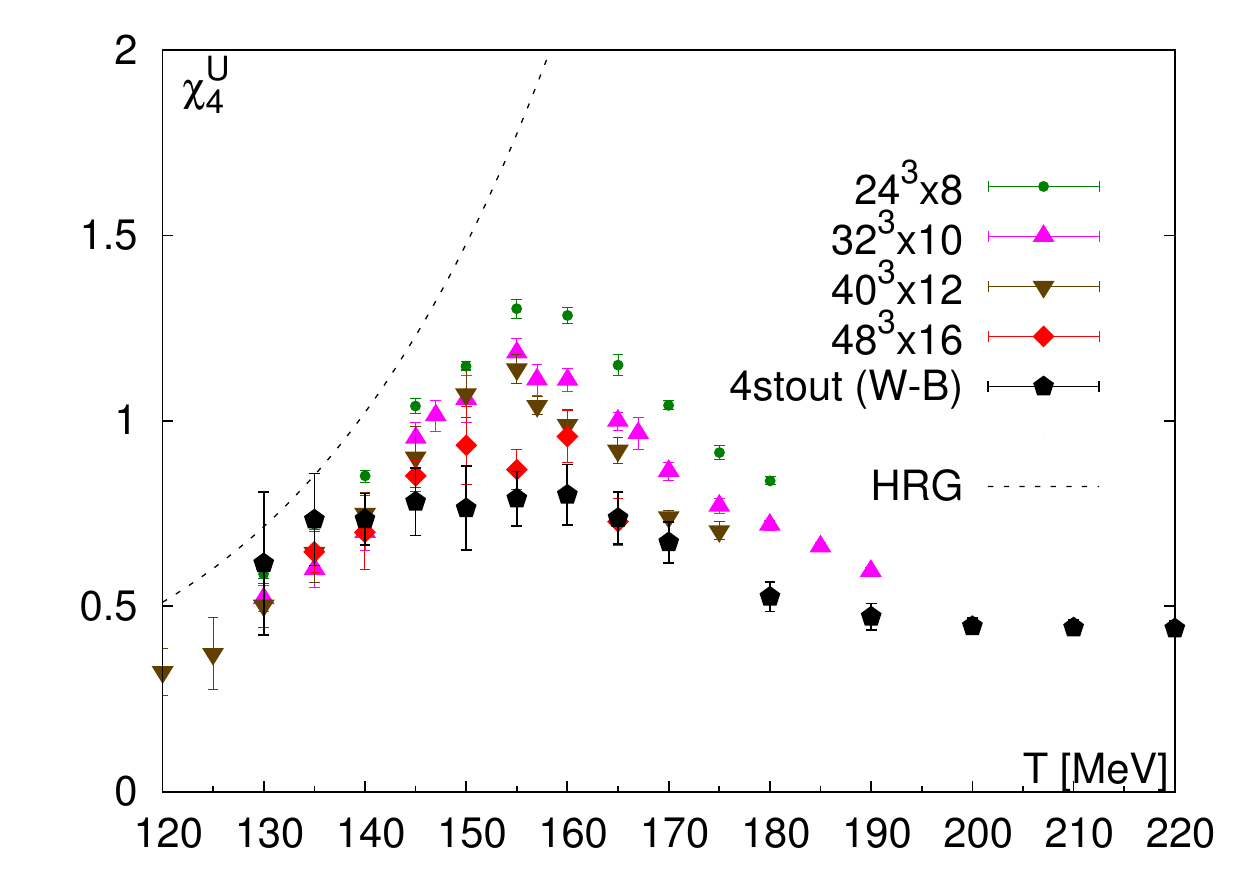}
\end{center}
\caption{\label{fig:c4U}
The single light quark number fourth order susceptibility in the cross-over region.
The extrapolation is driven by the $LT=3$ set of ensembles, which are
plotted together with the extrapolation and the HRG prediction.
(From $T>165~\mathrm{MeV}$ the continuum extrapolation uses only the $LT=4$
lattices (not shown)).  Since this is a potentially pion-driven observable and
the $N_t=24$ data are not sufficiently precise, the extrapolation is
based on $N_t=8\dots20$ lattices. In the cross-over region we consider this a
continuum estimate only. 
 }
\end{figure}

\begin{figure}[ht]
\begin{center}
\includegraphics[width=3in]{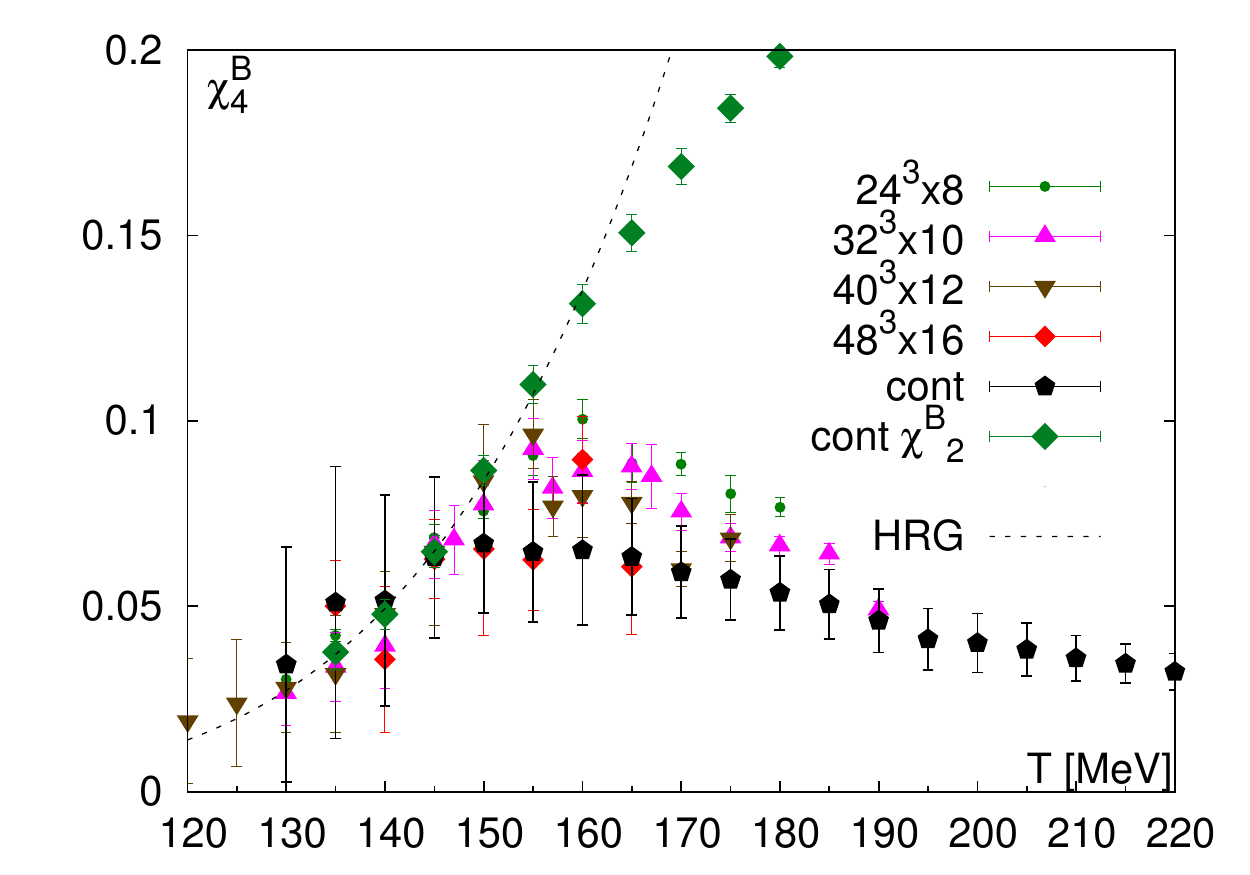}
\end{center}
\caption{\label{fig:c4B}
Continuum limit of the fourth moment of the baryon number distribution.
We also show the second moment. The HRG model gives the same result
for the two observables. The departure of $\chi^B_4$ from $\chi^B_2$
was interpreted as a signal of deconfinement in Ref.~\cite{Bazavov:2013dta}.
}
\end{figure}

Figures  \ref {fig:c4U} and \ref{fig:c4B} detail our continuum results for the fourth order cumulants.
The normalized strangeness \cite{Bellwied:2013cta} and baryon cumulants 
\cite{Borsanyi:2013hza,Bazavov:2013dta} have been published in earlier works.
Here we show the fourth derivative with respect to the light single quark
chemical potential (Fig.~\ref{fig:c4U}). On coarse lattices we see a strong
peak around the transition temperature. 

Such a peak has indeed been expected:
if QCD is in the chiral scaling regime with an O($N$) symmetry (i.e. the
light quark masses are small enough for QCD being nearly chiral) then this
scaling is expected to dominate the so called magnetic equation of state
\cite{Ejiri:2009ac}, which parametrizes the singular part of the free
energy as a reduced temperature and the quark masses that play the role
of the magnetic field in the O($N$) model's language. The chemical potential
enters through its shifting effect on the transition temperature. At finite
$\mu$, the reduced temperature is  $t\sim (T-T_c)/T_c - \kappa\mu^2/T^2$, where
$\kappa$ is the curvature of the QCD transition line
\cite{Kaczmarek:2011zz}. Using the critical exponents one has, for the
$n$-th derivative, a singular contribution of $\chi^B_n\sim |t|^{2-\alpha-n/2}$,
with $2-\alpha=\beta\delta(1+1/\delta)$ \cite{Friman:2011pf}. In the O(4)
universality class $\alpha=-0.2131(34)$ \cite{Engels:2003nq}.
The non-analytic contribution of $\chi^B_4(T)$ is thus singular in the chiral
limit and has a mild peak near $T_c$ at finite mass, while $\chi^B_6(T)$ changes
sign near $T_c$ \cite{Friman:2011pf}.

The data in Fig.~\ref{fig:c4U} show that the peak is strongly reduced on finer
lattices, as if we were moving away from the chiral limit. 
It will be interesting to see if this pattern is observed with other actions
with an improved dispersion relation.  Since here the $N_t=24$ data
have insufficient statistics, we cannot perform a controlled continuum
extrapolation at all temperatures: we call our result below $T_c$ a continuum
estimate.  What we see is that already at 145 MeV the Hadron Resonance Gas
model is unlikely to describe the lattice data. From our extrapolation based on
$N_t=8,10,12$ and $16$ lattices it is plausible to assume agreement at 135 MeV.

The baryon fourth moment shows milder lattice artefacts; here the large
statistical errors dominate over the systematic errors
(see Fig.~\ref{fig:c4B}). We also show $\chi^B_2(T)$ since the second and
fourth moment receive the same prediction from the HRG model, independently of how many
baryons and mesons are included in the resonance list. The point where $\chi^B_2(T)$ and
$\chi^B_4(T)$ are no longer consistent cannot be described by any resonance
list. Multi-baryon states are expected to lead to $\chi^B_4>\chi^B_2$, but here
we observe the opposite from $T>155$~MeV. The relation $\chi^B_4<\chi^B_2$ 
motivates the concept that the free energy is dominated by objects
with fractional baryon numbers: quarks.
Given the trend of the HRG model, it is conceivable that the departure point from 
the HRG model, and the respective maximum of the fourth-order derivative ($\chi^U_4$ or  $\chi^B_4$)
are very close in temperature.

\section{Results at high temperatures\label{sec:highT}}

\begin{figure}[ht]
\begin{center}
\includegraphics[width=3in]{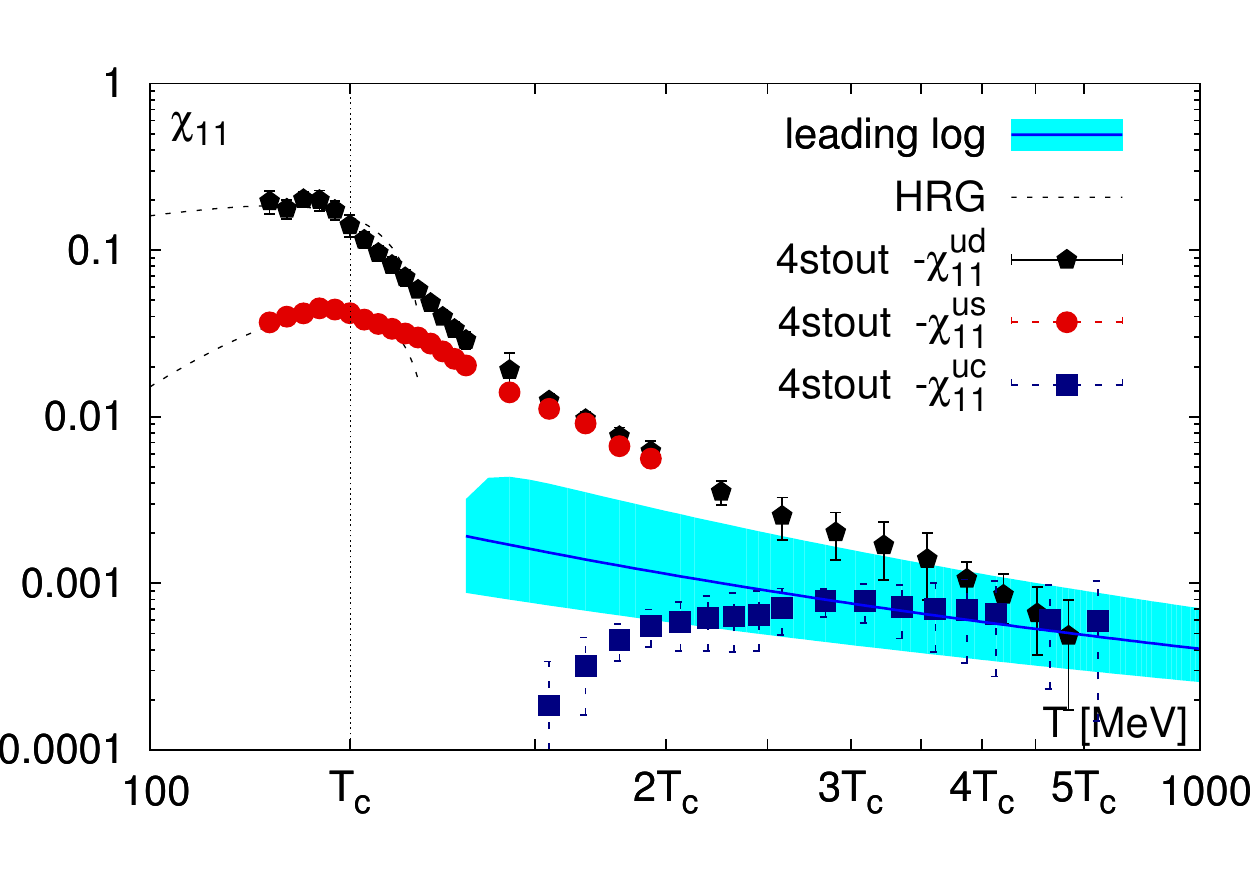}
\end{center}
\caption{\label{fig:c2ud_hot}
The off-diagonal quark number susceptibilities for various flavor combinations
 (see also Fig.~\ref{fig:c2ud}). The light correlator spans more than two
orders of magnitude in the temperature range between $T_c$ and $5T_c$ (using
the rescaling factor $T_c=155~\mathrm{MeV}$).  The leading
$\mathcal{O}(\alpha^3\log\alpha)$ perturbative result is from
Ref.~\cite{Blaizot:2001vr}. The mass of the strange quark becomes irrelevant
near $1.5T_c$. At $3T_c$ even the charm quark correlator agrees with the perturbative
result, even though the latter was calculated at zero mass.
}
\end{figure}

In this section we show our continuum extrapolated results at intermediate
and high temperatures. The first observables are the off-diagonal quark flavor
correlators, already shown in the transition region in Fig.~\ref{fig:c2ud}.
Increasing the temperature range (see Fig.~\ref{fig:c2ud_hot}), we actually see
that the value of the light-light correlator spans more than two orders
of magnitude between $T_c$ and $5T_c$. Between $4T_c$ and $5T_c$ the
leading perturbative log, which was calculated at zero quark mass
\cite{Blaizot:2001vr}, describes our data.  Our data suggest that the
light-charm correlator becomes compatible with the light-light correlator at about
$4T_c$, but its agreement with the leading log starts a bit earlier.
The mass of the strange quark is negligible in this observable already
at a temperature $\sim$240 MeV.

\begin{figure}[ht]
\begin{center}
\includegraphics[width=3in]{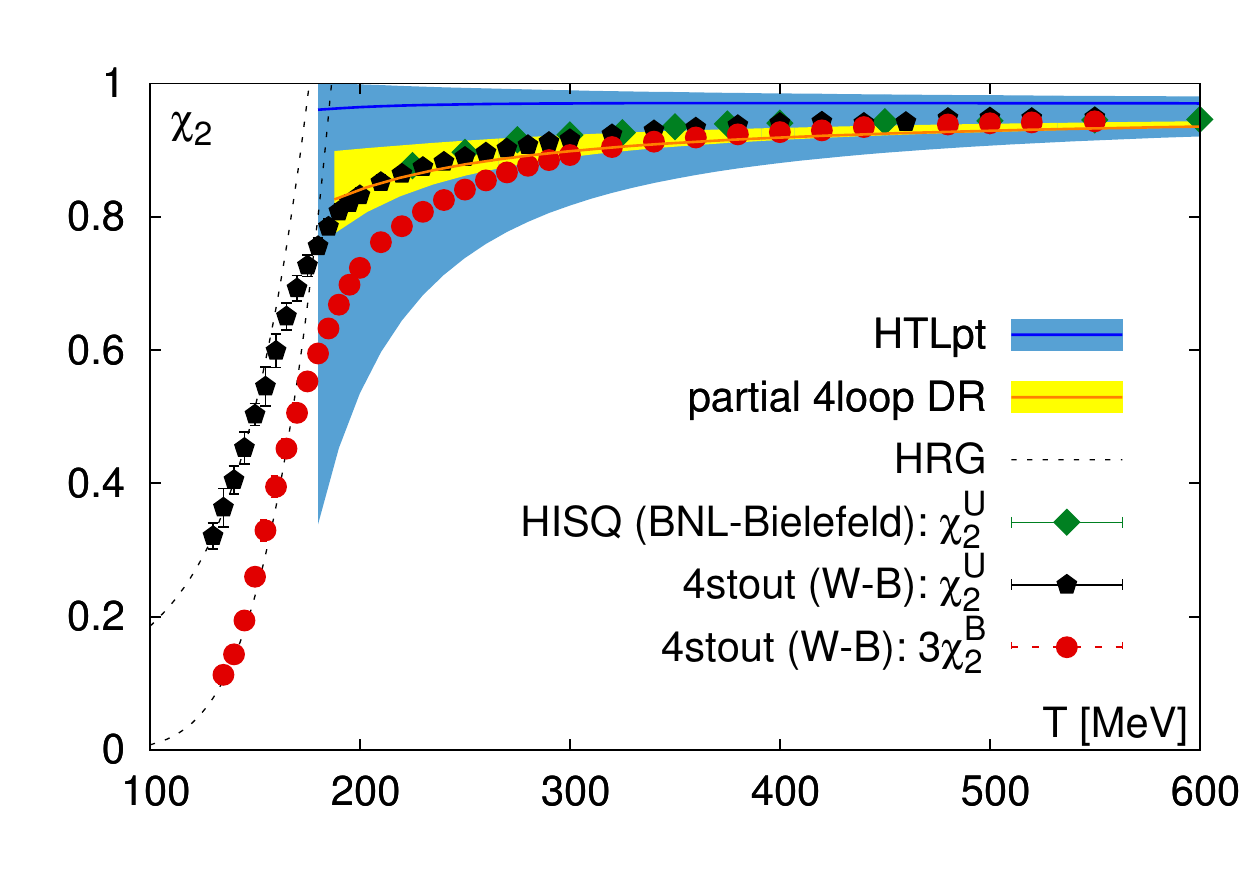}
\end{center}
\caption{\label{fig:c2U}
Second order diagonal fluctuations using the single quark chemical
potential $\chi^U_2$ vs. the baryon chemical potential $\chi^B_2$; we also
compare our data to the BNL-Bielefeld result \cite{Bazavov:2013uja}.
}
\end{figure}

For the light quark number susceptibility (Fig.~\ref{fig:c2U}) there are
continuum results available \cite{Borsanyi:2011sw,Bazavov:2013uja}. Here
we compare to the recent result with the HISQ action (with a combined analysis
also using p4 data) \cite{Bazavov:2013uja}. Our result is compatible with both Refs.~\cite{Borsanyi:2011sw,Bazavov:2013uja} within errorbars.
Here we also show the latest (improved) perturbative estimates,
based on hard thermal loops (HTL) \cite{Haque:2013sja} and dimensional
reduction (DR) \cite{Mogliacci:2013mca}. The improvement used 
in Ref.~\cite{Mogliacci:2013mca} has reduced the renormalization scale dependence
enormously. Our data are approximately one sigma higher than the upper edge
of the yellow band of the DR result. The central line of the band
is calculated at the renormalization scale $2\pi T$, the upper edge
at $4\pi T$ and the lower edge at $\pi T$.

\begin{figure}[ht]
\begin{center}
\includegraphics[width=3in]{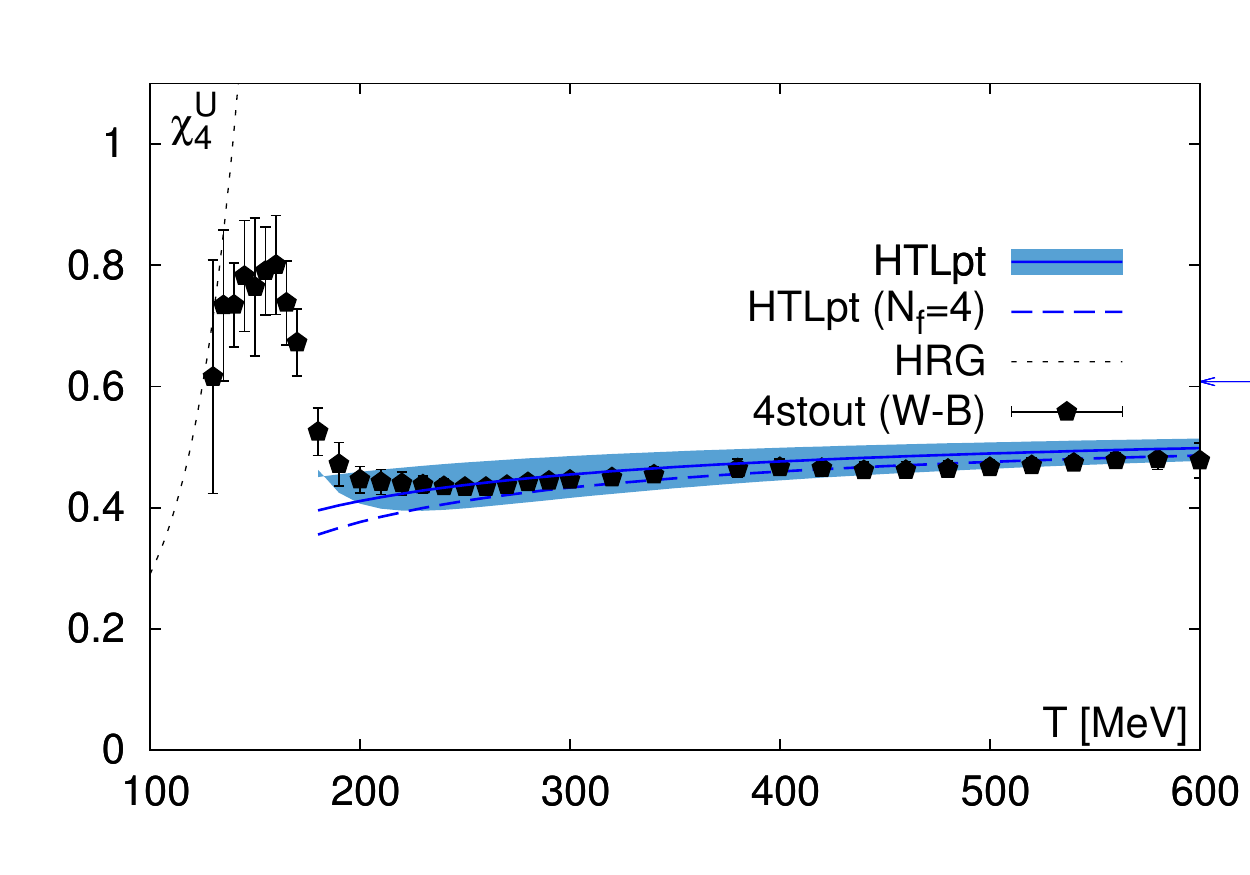}
\hfil
\includegraphics[width=3in]{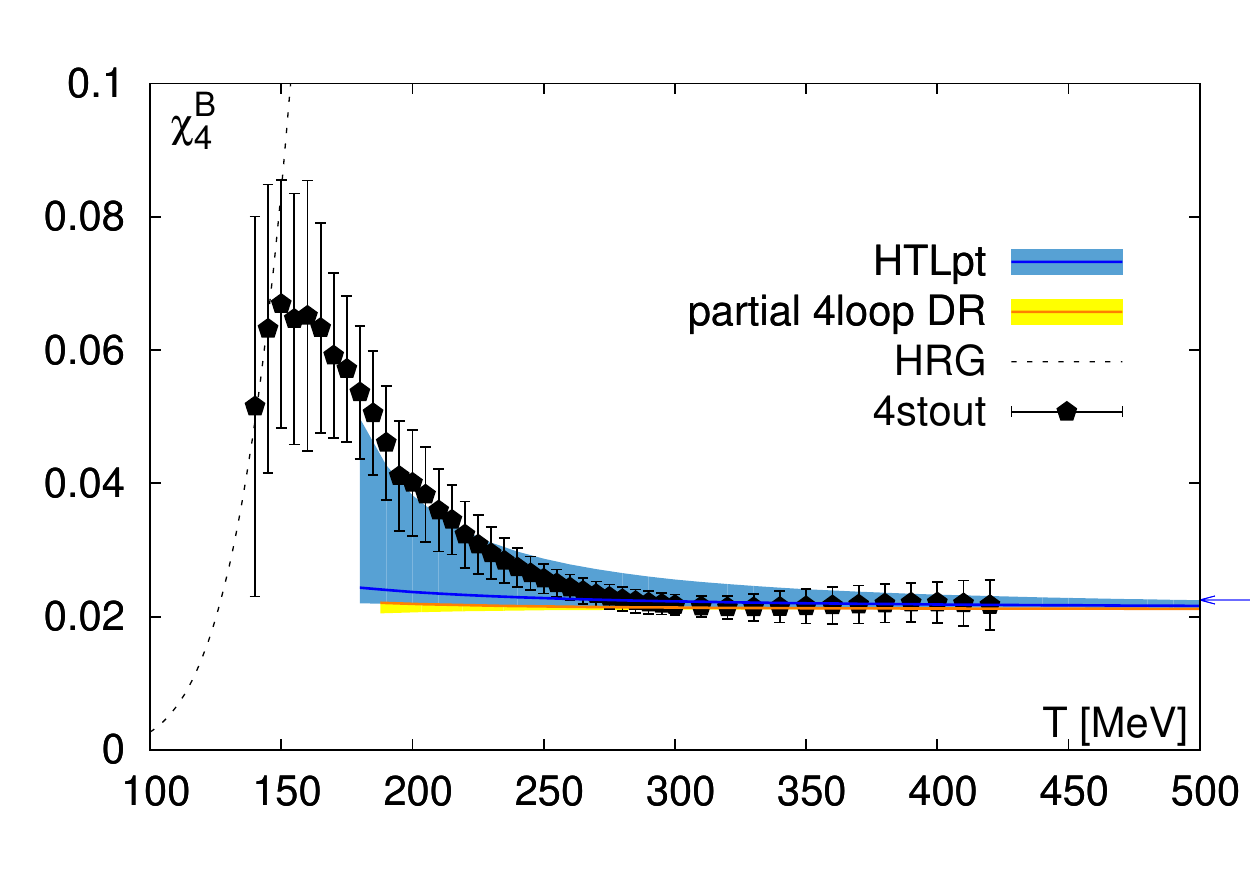}
\end{center}
\caption{\label{fig:c4}
Fourth order cumulants from our lattice study versus hard thermal loops
\cite{Haque:2014rua} and the result from  dimensional reduction (DR)
\cite{Mogliacci:2013mca}.
The small arrows on the right hand side mark the Stefan-Boltzmann limit.
}
\end{figure}

The fourth order cumulants at high and intermediate temperature are shown
in Fig.~\ref{fig:c4}. Both $\chi^U_4$ and $\chi^B_4$ are the fourth
derivative of the free energy with respect to the chemical potential, the
difference is that for the former the chemical potential is associated with
only one of the quarks, whereas for the latter it is associated with all
quarks at the same time.  Here the HTL results have a very small
renormalization scale dependence. 
The data confirms the HTL prediction that the Stefan-Boltzmann limit is
(almost) reached for $\chi^B_4$ at intermediate temperatures, $\chi^U_4$
approaches it much slower.  In both cases the improved and resummed
perturbative results give an accurate description of lattice data above 250 MeV.

This agreement may seem trivial since the lattice result is continuum
extrapolated and resummed perturbation theory is evaluated at high temperatures,
both approaches are expected to solve QCD. There is a subtle difference,
however, between HTL theory and lattice solutions.
We simulated our ensembles with physical quark masses and 2+1+1
dynamical flavors. HTL results, on the other hand, are available for
massless quarks only, and for $N_f=3$ as well as for $N_f=4$. The
mass of the strange quarks becomes irrelevant before we see agreement
between lattice data and HTL. At intermediate temperatures the large mass of
the charm makes the $N_f=3$ Hard Thermal Loop theory the closest match to our
setting. In order to compare the same observables we do not count the
baryon charge of the charm quark in $\chi^B_4$ and $\chi^B_2$.
To estimate the effect of the charm quark in the sea from the
improved perturbation theory side we plot the three-flavor and four-flavor 
result for $\chi^U_4$ together in Fig.~\ref{fig:c4} 
(see Ref.~\cite{Haque:2014rua}).

\begin{figure}[ht]
\begin{center}
\includegraphics[width=3in]{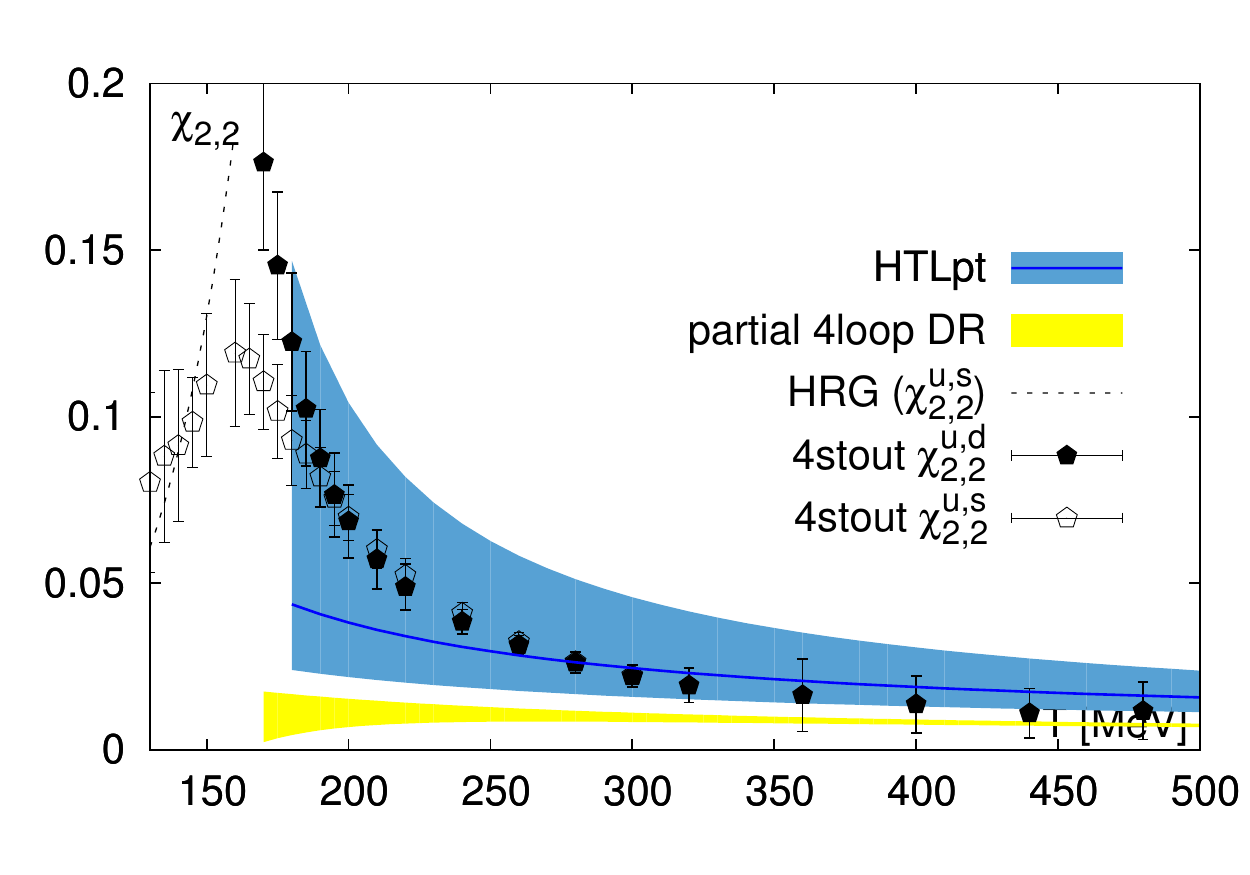}
\end{center}
\caption{\label{fig:UUDD}
The off-diagonal fourth order fluctuation at high temperature. Only this
off-diagonal derivative has a non-vanishing contribution in three-loop HTL 
\cite{Haque:2014rua}. The mass of the strange quark is irrelevant from approx.
200 MeV. Although the renormalization scale dependence between $\pi T$ and $4\pi T$
is large enough to contain the data, an agreement with the central line and
with its trend in temperature is reached at about 270 MeV.
The prediction of the DR method is also shown \cite{Mogliacci:2013mca}, there
is slight disagreement to HTL. Our data is compatible with both at high
temperature.
}
\end{figure}

We close our discussion with the off-diagonal fourth order correlator.
In Fig.~\ref{fig:UUDD} we show both the light-light and the light-strange
correlator. Here the effect of the strange quark mass diminishes even sooner,
at around 200 MeV.  The agreement with the HTL result starts at a temperature
$T\sim 250$ MeV, in accordance with the other observables
We also show the prediction of dimensional reduction \cite{Mogliacci:2013mca}.

\section{Concluding remarks\label{sec:concl}}

In this paper we introduced our thermodynamics program
with the four-level-smeared (4stout) staggered action. We focused on the fluctuations of conserved charges and updated our earlier result on second order fluctuations
\cite{Borsanyi:2011sw}. Since our first paper on fluctuations, we have introduced
very fine lattices ($N_t=24$) in the transition range and extended the analysis to
high temperatures where a comparison to resummed and improved perturbation theory
is possible. We have also presented diagonal and off-diagonal fourth order cumulants.
Here our data could be used to determine the lowest temperature for the
three-loop HTL approximation: approx. 250 MeV.

We have studied whether the hadron resonance gas (HRG) model gives an adequate
description of the fluctuation data. We find that well below the deconfinement
temperature, i.e. around 130 MeV, all studied observables are well described by the HRG model.
This was the most difficult to demonstrate for the
fourth moment of the net charge distribution $\chi^4_Q$, which is a candidate
for the freeze-out thermometer at the LHC. In this case, after adding a
$96^3\times 32$ lattice to the analysis ($a=0.047$~fm), our continuum extrapolation based on
 $N_t=20,~24$ and 32 lattices is consistent with the HRG model prediction.

It is very likely that HRG does not describe \textit{all} aspects of 
fluctuations in QCD thermodynamics below the transition. But for quantities
for which it does one can introduce the highest temperature of agreement between
lattice and HRG. This indicator of deconfinement is unavoidably model-dependent,
even if one considers combinations that do not or only weakly depend on the
actual list of resonances. This temperature can, however, be determined as long
as the continuum limits are feasible with a sufficient precision. The data
on our plots show in most cases an agreement up to $\sim T_c$, which can move
to a lower temperature as our precision improves. This should not be confused with
the limiting temperature of the Hagedorn spectrum, which can be higher. 
The temperature of highest agreement is not the same for all fluctuations
as it was also suggested in Ref.~\cite{Bellwied:2013cta}, e.g. $\chi^U_4$
and very possibly $\chi^Q_4$ depart from the HRG estimates at lower
temperatures. This may be a signal of the limitations of the HRG approach, but
also suggests that the transition is a broad cross-over.

\section{Acknowledgements} 
S. B. acknowledges the valuable discussions, correspondence and data exchange
with Najmul Haque, Silvain Mogliacci, Mike Strickland, Nan Su and 
Aleksi Vuorinen.

This project was funded by the DFG grant SFB/TR55.  S. D. Katz is funded by
the "Lend\"ulet" program of the Hungarian Academy of Sciences
((LP2012-44/2012).  The work of R.  Bellwied is supported through DOE grant
DEFG02-07ER41521. Our code also uses the software package of
Ref.~\cite{Hernandez:2005:SSF}. An award of computer time was provided by the INCITE program. This research used resources of the Argonne Leadership Computing Facility, which is a DOE Office of Science User Facility supported under Contract DE-AC02-06CH11357. This research also used resources of
the PRACE Research Infrastructure resource JUQUEEN at FZ-J\"ulich, Germany.
The authors gratefully acknowledge the Gauss Centre for Supercomputing (GCS) 
for providing computing time for a GCS Large-Scale Project on the GCS share of 
the supercomputer JUQUEEN \cite{juqueen} at J\"ulich Supercomputing Centre (JSC). 
For this research we also used the QPACE machine supported by the Deutsche
Forschungsgesellschaft through the research program SFB TR55, and the GPU
cluster at the Wuppertal University.
\bibliography{thermo}{}
\end{document}